\documentclass[a4paper]{elsarticle}

\usepackage[colorlinks,bookmarks=false]{hyperref}

\usepackage[latin1]{inputenc}
\usepackage[T1]{fontenc}
\usepackage[english]{babel}
\addto\captionsenglish{}

\usepackage{adjustbox}
\usepackage{booktabs}
\usepackage{multicol}
\newcommand{\specialcell}[2][l]{\begin{tabular}[#1]{@{}l@{}}#2\end{tabular}}

\usepackage{amssymb}
\usepackage{amsmath}
\usepackage{amsthm}
\usepackage{bm}
\usepackage[squaren,Gray]{SIunits}

\usepackage{subfig}
\graphicspath{{./pictures/}}

\usepackage{tikz}
\usepackage{pgf}
\usepackage{pgfplots}
\pgfplotsset{compat=1.14}
\usetikzlibrary{plotmarks}
\usetikzlibrary{shapes,arrows,backgrounds,patterns,calc}
\makeatletter
\def\input@path{{./tikz/}}
\makeatother

\numberwithin{equation}{section}

\journal{Computers and Fluids}

\begin{document}
	
\begin{frontmatter}
			
\title{Simulations of intermittent two-phase flows in pipes using smoothed particle hydrodynamics}
				
\author[1,3]{Thomas Douillet-Grellier} \corref{cor1}
\author[2]{Florian De Vuyst} 
\author[3]{Henri Calandra}
\author[3]{Philippe Ricoux}

\address[1]{CMLA, ENS Cachan, CNRS, Universit\'e Paris-Saclay, 94235 Cachan, France}
\address[2]{LMAC, EA 2222, Sorbonne Universit\'es, Universit\'e de Technologie de Compi\`egne, 60200 Compi\`egne, France}
\address[3]{Total S.A., Tour Coupole, 92078 Paris La D\'efense Cedex, France}
		
\cortext[cor1]{thomas.douillet-grellier@ens-paris-saclay.fr}
		
\begin{abstract}
Slug flows are a typical intermittent two-phase flow pattern that can occur in submarine pipelines connecting the wells to the production facility and that is known to cause undesired consequences. In this context, computational fluid dynamics appears to be the tool of choice to understand their formation. However, few direct numerical simulations of slug flows are available in the literature, especially using meshless methods which are known to be capable of handling complex problems involving interfaces.

In this work, a 2D study of the instability processes leading to the formation of intermittent flows in pipes is conducted using an existing multiphase smoothed particle hydrodynamics formulation associated with inlet and outlet boundary conditions. This paper aims to demonstrate the applicability of smoothed particle hydrodynamics to a given set of close-to-industry cases.

First, we check the ability of our implementation to reproduce flow regimes predicted by Taitel and Duckler's flow map. Then, we focus on the transition processes from one flow pattern to the other. Finally, we present the results obtained for more realistic cases with high density and viscosity ratios.
			
\end{abstract}
		
		
\end{frontmatter}

\section{Introduction}

Two-phase flows problems involving non-miscible fluids with a deformable interface are encountered in numerous industrial and scientific applications~\cite{Hewitt}. One of the main examples is the transportation of oil and gas (or oil and water) through pipelines in the petroleum industry. However, gas-liquid flows are also present in chemical and heat transfer systems such as evaporators, boilers, condensers, distillation processes, air-conditioning or refrigerators and liquid-liquid flows are fundamental in solvent extraction equipments such as pulsed columns. Furthermore, all these two phase flows can also be found in natural environments and are studied in meteorology for instance. 

The main focus of this work is the emergence of intermittent flow patterns (also known as slug or plug flow~\cite{Fabre1992}) which only occurs in gas-liquid or liquid-liquid flows. In particular, in pipeline networks, these patterns are highly undesirable~\cite{Sausen2012}. Those slug patterns, that can measure up to tens of meters, are known to damage facilities (separators flooding, compressors starving, water hammer phenomenon) and to reduce flow efficiency. Besides, intermittent flows are also commonly found in microfluidic applications in the chemical industry where they can improve reactions performances~\cite{Triplett1999,yao2017}.

The mechanisms of generation of slug flows are well known. On one hand, in the case of horizontal pipes, hydrodynamic slugs are induced by the Kelvin-Helmholtz instability. Under certain velocity and geometry conditions, liquid waves can grow in the flow causing a local increase of the gas velocity and a local decrease of the pressure. Consequently, a suction force starts to move the interface higher until reaching the top of the pipe and forming a slug. A common way to avoid the appearance of hydrodynamic slugs is to use flow regime maps to operate with flow conditions that are unlikely to generate them.

On the other hand, intermittent flows can be caused by the combined effects of gravity and terrain geometry on which the pipe lies~\cite{SnchezSilva2013}. The path of the pipe can have low spots, like elbows, in which the liquid is trapped. It accumulates until reaching the top of the pipe and it is then carried away by the flow forming a slug. Alternatively, severe slugging can occur in risers. In a nutshell, a liquid slug begins to form at the riser base blocking the incoming gas. Moved by the increasing gas pressure, the liquid fills up the riser forming a slug that flood the separator at the end of the riser. Additionally, slugging can be generated by a flow rate change or by pigging.

The main question when studying slugging is to know whether or not it will occur and eventually to find the criterion that triggers its formation~\cite{zhang2017}. If it does, the quantities of interest are the size of slugs and their transit time and frequency.

The understanding of the formation of intermittent flow patterns has been a lively research area for years. In this context, Computational Fluid Dynamics (CFD) softwares~\cite{lu2015experimental,pedersen2017} emerged as a useful tool to predict the appearance of a slug flow regime in the oil and gas industry. 

Early simulations were based on steady state models~\cite{taitel1980,viggiani1988} and transient models~\cite{sarica1991} and were able to simulate slugging in gravity-dominated flows. In the oil and gas industry, two commercial softwares are competing for slugging simulation : OLGA developed by SPT group~\cite{bendiksen1991} and LedaFlow, proposed by Kongsberg~\cite{kongsberg2014}. A detailed comparison of both softwares that can be found in~\cite{belt2011} concludes that although performing equally well on simple cases, they have trouble to simulate complex cases with a dominant gas phase.

From an academic perspective, different methods have been used for slug flow modeling such as volume-of-fluid~\cite{Taha2004, al2016numerical}, level-set~\cite{Fukagata2007,lizarraga2016study}, lattice boltzmann~\cite{Yu2007} or phase field~\cite{He2008,xie2017} but they are mostly focused on microfluidic problems.

The method known as Smoothed Particle Hydrodynamics (SPH) is becoming more and more popular throughout the years. Indeed, thanks to its meshless nature, SPH emerged as an interesting tool to model multiphase flows and many authors have applied it to solve complex multiphase problems~\cite{Wang2016}. Nevertheless, SPH or more generally meshless simulations of slug flows are not common in the literature despite the inherent benefits of being meshfree for tackling interfaces problems. To the best of our knowledge, the only previous SPH study on the topic is available in~\cite{minier2016} and is focused on the transition from bubbly flow to slug flow using periodic boundary conditions and a gravity-based driving force.

Our goal with this work is to contribute to the ongoing effort to show that SPH can be applied to industrial cases. To this end, we used recent multiphase SPH features available in the literature to a set of close-to-industry test cases.

In this paper, we first detail the multiphase SPH formulation introduced in~\cite{Hu2006} including the surface tension model presented in~\cite{lafaurie1994} and with particular emphasis on how to handle inlet/outlet boundary conditions in a multiphase context in sections~\ref{governing} to~\ref{sec:bc}. Then, in sections~\ref{valid} and~\ref{trans}, we verify the ability of our SPH model to recover different flow regimes predicted by Taitel and Dukler's flow map~\cite{Taitel1976} and we study the transition processes between two flow patterns. Finally, in section~\ref{oilgas}, we present two more realistic cases involving high density and viscosity ratios.

A collection of quantitative validation cases of our implementation for both single phase and multiphase problems is presented in~\ref{appendix}.

\section{Governing Equations}\label{governing}

In this section, the main governing equations of the considered problem are recalled in details.
 
\subsection{Single phase balance equations}
The governing equations of the problem for a single fluid phase consist of mass and momentum conservation equations in a Lagrangian system, and are given as

\begin{align}
	\frac{D \rho}{D t} &= - \rho \nabla \cdot \bm{u}, \label{continuity_equation}\\ 
	\rho \frac{D \bm{u}}{D t} &= -\nabla p + \nabla \cdot \bm{\tau} + \rho \bm{g},\label{momentum_equation}
\end{align}

\noindent with $\bm{u}$ fluid velocity, $\rho$ fluid density, $\bm{\tau}$ viscous stress tensor, $p$ fluid pressure, $\mu$ fluid dynamic viscosity, $\bm{g}$ gravity and $D/Dt$ denotes the material derivative following the motion. Note that, in the case of an incompressible fluid with a constant viscosity, the viscous term reduces to $\nabla \cdot \bm{\tau} = \mu \nabla^2 \bm{u}$ with $\mu$ the fluid viscosity.

A constitutive relation for the evaluation of $p$ has to be added to the governing equations \eqref{continuity_equation}-\eqref{momentum_equation} to close the system. In this paper, the Tait's equation of state has been used. It reads

\begin{equation}\label{tait_eos}
p = \frac{c^2 \rho^0}{\gamma} \left[ \left( \frac{\rho}{\rho_0} \right)^{\gamma} -1 \right] + p_0,
\end{equation}

\noindent with $c$ fluid speed of sound (here constant), $\gamma$ fluid adiabatic index, $\rho_0$ fluid initial density and $p_0$ background pressure.

This approach is known as the Weakly Compressible SPH formulation (WCSPH). It is not a truly incompressible approach since the density is allowed to vary because of the Lagrangian nature of the algorithm. This artificial compressibility has to be as weak as possible and is controlled by the speed of sound $c$. Universal guidelines to set a good value for $c$ and $p_0$ are not available. It is common to set a value and adjust it on a case by case basis. In this paper, given a reference length $L_{\text{ref}}$ and a reference speed $U_{\text{ref}}$, the following formulas, taken from~\cite{morris1999surfacetension}, were used

\begin{equation}\label{sspb}
\left\{ 
\begin{array}{l l}
	c_\alpha &= \max\left( \frac{U_{\text{ref}}}{\sqrt{\Delta \rho}},\sqrt{\frac{\lvert \bm{g} \rvert L_{\text{ref}}}{\Delta \rho}}, \sqrt{\frac{\sigma^{\alpha\beta}}{\rho_{0\alpha} L_{\text{ref}}}}, \sqrt{\frac{\mu_\alpha U_{\text{ref}}}{\rho_{0\alpha} L_{\text{ref}} \Delta \rho}} \right),\quad \forall \alpha \in \{1,\ldots,N_{\text{phases}}\},\\
	p_0 &= \max_{\alpha \in \{1,\ldots,N_{\text{phases}}\}} \frac{c^2_\alpha \rho_{0\alpha}}{\gamma_\alpha},
 \end{array} \right.
\end{equation}

\noindent with $\Delta \rho = 0.01$ to enforce (not strictly) a maximum variation of $1\%$ of the density field and $\sigma^{\alpha \beta}$ the surface tension coefficient between phase $\alpha$ and $\beta$. $N_{\text{phases}}$ is the number of different phases. Some authors recommend to use $p_0 = 0.1 \max_{\alpha \in \{1,\ldots,N_{\text{phases}}\}} \frac{c^2_\alpha \rho_{0\alpha}}{\gamma_\alpha}$ but we were not able to obtain stable simulations with such low values. The background pressure helps stabilizing the simulations but can lead to pressure noises if set too high~\cite{violeau2014timestep}. In this work, we have not done an exhaustive sensitivity study and used the expression given in~\eqref{sspb}. It might be possible to adjust more carefully the background pressure for each case to attenuate pressure oscillations.

\subsection{Surface tension}\label{sec:st}

In the case of a multiphase flow system, each phase is governed by the set of equations \eqref{continuity_equation}, \eqref{momentum_equation} and \eqref{tait_eos}. An interaction force has to be added to the momentum equation to model surface tension between fluids. The continuum surface stress method introduced by \cite{lafaurie1994} has been used to that end. In this approach, the surface tension force per unit volume is expressed as the divergence of the capillary pressure tensor

\begin{equation}
		\bm{F}^{st} = -\nabla \cdot \Pi,
\end{equation}

\noindent with $\Pi$ the capillary pressure tensor defined by

\begin{equation}\label{st_pi}
\Pi = \sum_{\alpha,\beta \mid \alpha < \beta} \Pi^{\alpha \beta},
\end{equation}

\noindent where $\alpha, \beta \in \{1,\ldots,N_{\text{phases}}\}$ and $\Pi^{\alpha \beta}$ is expressed as

\begin{equation}\label{st_piab}
\Pi^{\alpha \beta} = -\sigma^{\alpha \beta} \left( \bm{I} - \tilde{\bm{n}}^{\alpha \beta} \otimes \tilde{\bm{n}}^{\alpha \beta} \right) \delta^{\alpha \beta},
\end{equation}

\noindent with $\tilde{\bm{n}}^{\alpha \beta}$ the unit normal vector from phase $\alpha$ to phase $\beta$, $\sigma^{\alpha \beta}$ the surface tension coefficient between phase $\alpha$ and phase $\beta$, $\delta^{\alpha \beta}$ a well-chosen surface delta function and $\bm{I}$ the identity matrix.

\begin{figure}[bthp]
	\centering
	\resizebox{0.65\textwidth}{!}{\input{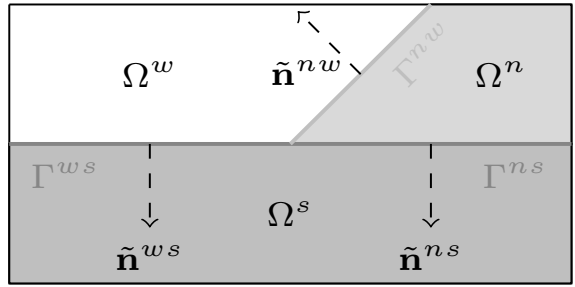}}
	\caption{A triple point contact line between a non wetting phase $\Omega_n$, a wetting phase $\Omega_w$ and a solid phase $\Omega_s$ with their associated unit normal vectors $\tilde{\bm{n}}_{ab}$ and boundary lines $\Gamma_{ab}$}
	\label{triple_point}
\end{figure}

In the case of a three-phase system with a wetting phase $s$, a non wetting phase $n$ and a solid phase $s$ as described in Figure~\ref{triple_point}, the stress tensor reads $\Pi = \Pi^{ns} + \Pi^{ws} + \Pi^{nw}$.

There are at least two other approaches to represent surface tension in an SPH formulation. One is very similar to the one that was used in this paper and is based on curvature computation \cite{brackbill1992,morris1999surfacetension} and the other one is based on pairwise forces~\cite{Tartakovsky20161119}.

The full set of governing equations including surface tension for an incompressible fluid is given as

\begin{equation}
\left\{ 
\begin{array}{l l}
 \frac{D \rho}{D t} &= - \rho \nabla \cdot \bm{u},\\ 
 \frac{D \bm{u}}{D t} &= -\frac{\nabla p}{\rho} + \nu \nabla^2 \bm{u}  +\frac{\bm{F}^{st}}{\rho}+  \bm{g},\\
p &= \frac{c^2 \rho_0}{\gamma} \left[ \left( \frac{\rho}{\rho_0} \right)^{\gamma} -1 \right] + p_0, \end{array} \right.
\end{equation}

\noindent with $\nu=\frac{\mu}{\rho}$ the kinematic viscosity.

\section{SPH Formulation}\label{sphform}

In this section, we introduce the SPH formulation used throughout this paper.

\subsection{Standard SPH Formulation}

An SPH discretization consists of a set of points with fixed volume, which possess material properties and interact with all neighboring particles through a weighting function (or smoothing kernel)~\cite{gingold1977smoothed}. A particle's support domain, $\Lambda$, is given by its smoothing length, $h$, which is the radius of the smoothing kernel. In all simulations presented in this paper, $h=2\Delta r$ where $\Delta r$ is the initial particle spacing. To obtain the value of a function at a given particle location, values of that function are found by taking a weighted (by the smoothing function) interpolation from all particles within the given particle's support domain. Analytical differentiation of the smoothing kernel is used to find gradients of this function. Detailed concepts and descriptions of this method are given by Monaghan~\cite{monaghan1992smoothed}.

To begin, we define the following kernel estimation

\begin{equation}\label{eq:4.1}
\bm{A}(\bm{x})=\int_{\Omega}{\bm{A}(\bm{x}')W(\bm{x-x'},h)}\, d \bm{x}',~~~~~~~~\forall~\bm{x}
\in\Omega\subset\mathbb{R}^{d}~~~~~~
\end{equation}

\noindent where $\bm{A}$ is a vector function of the position vector $\bm{x}$, $\Omega$ is the integral space containing the point $\bm{x}$, and $W(\bm{x-x'},h)$ is the smoothing kernel. The interpolated value of a function $\bm{A}$ at the position $\bm{x}_a$ of particle $a$ can be expressed using SPH smoothing as

\begin{equation}\label{eq:4.2}
\bm{A}(\bm{x}_a)=\sum_{b\in\Lambda_{a}}\bm{A}_{b}\frac{m_b}{\rho_{b}}W(\bm{x}_a-\bm{x}_{b},h),
\end{equation}

\noindent in which $\bm{A}_{b}=\bm{A}(\bm{x}_b)$, $m_b$ and $\rho_{b}$ are the mass and the density of neighboring particle $b$. The set of particles $\Lambda_{a}=\{ b\in \mathbb{N}\mid \lvert\bm{x}_{a}-\bm{x}_{b}\rvert \leq \kappa h \}$ contains neighbors of particle $a$ and lie within its defined support domain. The coefficient $\kappa$ depends on the choice of the kernel, it is equal to $2$ for the $5^{th}$ order $\mathcal{C}^2$ Wendland kernel function~\cite{Wendland1995,Dehnen11092012}) used in this paper. In 2D, this kernel is expressed as follows

\begin{equation}
		W(q,h) = \frac{7}{4\pi h^2} \left(1-\frac{q}{2}\right)^4 \left(1 +2q\right),
\end{equation}

\noindent with $q=\frac{\lvert \bm{x}_a-\bm{x}_{b} \rvert}{h}$, $q \leq 2$.

Gradient and divergence operators of the function $\bm{A}$ at the position of particle $a$ are evaluated by differentiating the smoothing kernel $W$ in equation~\eqref{eq:4.2} as

\begin{align}\label{eq:4.3}
\nabla \bm{A}(\bm{x}_a)&=\sum_{b\in\Lambda_{a}}\bm{A}_{b}\frac{m_b}{\rho_{b}}\nabla_a W(\bm{x}_a-\bm{x}_{b},h),\\
\nabla \cdot \bm{A}(\bm{x}_a)&=\sum_{b\in\Lambda_{a}}\bm{A}_{b}\frac{m_b}{\rho_{b}} \cdot\nabla_a  W(\bm{x}_a-\bm{x}_{b},h).
\end{align}

\noindent In practice, Libersky \textit{et al.} find that by exploiting the symmetric properties of the kernel, a more accurate formulation is found~\cite{libersky1993high} as

\begin{align}\label{eq:4.3b}
\nabla \bm{A}(\bm{x}_a)&= -\frac{1}{\rho_{a}} \sum_{b\in\Lambda_{a}}\bm{A}_{ab} m_b\nabla_a W_{ab},\\
\nabla \cdot \bm{A}(\bm{x}_a)&= -\frac{1}{\rho_{a}} \sum_{b\in\Lambda_{a}}\bm{A}_{ab} m_b \cdot\nabla_a W_{ab}.
\end{align}

\noindent where, for the sake of clarity, $W(\bm{x}_a-\bm{x}_{b},h)$ and $\bm{A}_{a}-\bm{A}_{b}$ have been denoted $W_{ab}$ and $\bm{A}_{ab}$ respectively. These notations will be used in the rest of the paper.

\subsection{Multiphase SPH Formulation}

Since the introduction of the SPH method, several specific formulations designed for multiphase problems have been introduced. Among them, one can mention \cite{colagrossi2003, grenier2009, tofighi2013} and \cite{zainali2013}. In this paper, the formalism introduced in \cite{Hu2006} has been used. In this framework, the continuity equation~\eqref{continuity_equation} is not discretized directly. A common practice in SPH for flows that do not involve any free surface, the density is directly evaluated through a kernel summation. This approach gives an exact solution to the continuity equation. Traditionally in SPH, this kernel summation is $\rho_{a} = \sum_{b\in\Lambda_{a}} m_{b} W_{ab}$ but in the present framework, it is 

\begin{equation}
	\rho_{a} = m_{a}\sum_{b\in\Lambda_{a}} W_{ab}.
\end{equation}
\noindent The difference is that the density evaluation for a given particle $a$ does not take into account the masses of neighboring particles which allows the treatment of density discontinuities.

Discretized gradient and divergence operators in this formalism are given by

\begin{align}
	\nabla \bm{A}(\bm{x}_a) =  \sum_{b\in\Lambda_{a}} \left(\frac{\bm{A}_{a}}{\Theta_a^2} + \frac{\bm{A}_{b}}{\Theta_b^2}\right) \Theta_a \nabla_a W_{ab},\\
	\nabla \cdot \bm{A}(\bm{x}_a) =  \sum_{b\in\Lambda_{a}} \left(\frac{\bm{A}_{a}}{\Theta_a^2} + \frac{\bm{A}_{b}}{\Theta_b^2}\right) \Theta_a \nabla_a \cdot W_{ab},
\end{align}

\noindent where $\Theta_a=\frac{\rho_{a}}{m_{a}}$. It follows that the discretized pressure term $-\frac{\nabla p_{a}}{\rho_a}$ for a particle $a$ is given by

\begin{align}
-\frac{\nabla p_{a}}{\rho_a} = -\frac{1}{m_a} \sum_{b\in\Lambda_{a}} \left(\frac{p_{a}}{\Theta_a^2} + \frac{p_{b}}{\Theta_b^2}\right) \nabla_a W_{ab},
\end{align}

\noindent where the pressures $p_a$ and $p_b$ are computed using the equation of state~\eqref{tait_eos}. The discretized viscous term $\nu_a \nabla^2 \bm{u}_{a}$ for a particle $a$ is discretized using the inter-particle averaged shear stress~\cite{flekkoy2000} leading to

\begin{align}
\nu_a \nabla^2 \bm{u}_{a} = \frac{1}{m_a} \sum_{b\in\Lambda_{a}} \frac{2 \mu_a \mu_b}{\mu_a + \mu_b} \left(\frac{1}{\Theta_a^2} + \frac{1}{\Theta_b^2}\right) \frac{\bm{x}_{ab} \cdot \nabla_a W_{ab}}{\lvert \bm{x}_{ab} \rvert^2 + \eta^2} \bm{u}_{ab},
\end{align}

\noindent where $\eta = 0.01 h$ is a safety factor to avoid a division by zero. The discretized surface tension term $\frac{\bm{F}^{st}_a}{\rho_{a}}$ for a particle $a$ is given by

\begin{align}
\frac{\bm{F}^{st}_a}{\rho_{a}} = -\frac{1}{m_a} \sum_{b\in\Lambda_{a}} \left(\frac{\Pi_{a}}{\Theta_a^2} + \frac{\Pi_{b}}{\Theta_b^2}\right) \nabla_a W_{ab},
\end{align}

\noindent where the stress tensors $\Pi_{a}$ and $\Pi_{b}$ are computed using equations~\eqref{st_pi}-\eqref{st_piab}. 

Note that in SPH, the evaluation of normals is performed through the computation of the gradient of a color function $\chi$ defined for a given particle $a$ and a given phase $\alpha$ as

\begin{align}
\chi^{\alpha}_a =
\begin{cases}
1\quad\text{if }a\in\text{phase }\alpha,\\
0\quad\text{else}.
\end{cases}
\end{align}

\noindent The normal vector $\bm{n}^{\alpha \beta}_a$ of particle $a$ belonging to phase $\alpha$ to the interface $\alpha\beta$ is then computed doing

\begin{equation}
	\bm{n}^{\alpha \beta}_a = \nabla \chi^{\alpha \beta}_a = \sum_{b\in\Lambda_{a}} \left(\frac{\bm{\chi}^{\beta}_{a}}{\Theta_a^2} + \frac{\bm{\chi}^{\beta}_{b}}{\Theta_b^2}\right) \Theta_a \nabla_a W_{ab}.
\end{equation}

\noindent The surface delta function $\delta^{\alpha\beta}_a$ is chosen to be equal to $\lvert \bm{n}^{\alpha \beta}_a \rvert$ and $\tilde{\bm{n}}^{\alpha \beta} = \bm{n}^{\alpha \beta}_a / \lvert \bm{n}^{\alpha \beta}_a \rvert$.

Moreover, due to its Lagrangian nature, the SPH particles are moved using simply

\begin{equation}
	\frac{D \bm{x}_a}{Dt} = \bm{u}_a.
\end{equation}

Ultimately, the full multiphase SPH formulation for a particle $a$ can be summarized as follows

\begin{equation}\label{fullformulation}
\left\{ 
\begin{array}{l l}
	\rho_{a} &= m_{a}\sum_{b\in\Lambda_{a}} W_{ab}, \\
	\frac{D \bm{u}}{D t} &= -\frac{1}{m_a} \sum_{b\in\Lambda_{a}} \left(\frac{p_{a}\bm{I}+\Pi_{a}}{\Theta_a^2} + \frac{p_{b}{I}\bm+\Pi_{b}}{\Theta_b^2}\right) \nabla_a W_{ab}\\
	&\quad+ \frac{1}{m_a} \sum_{b\in\Lambda_{a}} \frac{2 \mu_a \mu_b}{\mu_a + \mu_b} \left(\frac{1}{\Theta_a^2} + \frac{1}{\Theta_b^2}\right) \frac{\bm{x}_{ab} \cdot \nabla_a W_{ab}}{\lvert \bm{x}_{ab} \rvert^2 + \eta^2} \bm{u}_{ab}\\
	&\quad+ \bm{g},\\
	p_a &= \frac{c^2_a \rho_{0a}}{\gamma_a} \left[ \left( \frac{\rho_a}{\rho_{0a}} \right)^{\gamma_a} -1 \right] + p_0,\\
	\frac{D \bm{x}_a}{Dt} &= \bm{u}_a. \end{array} \right.
\end{equation}

\subsection{Corrective terms}
In this work, three SPH correction procedures have been used. First, as suggested in~\cite{bonet1999}, the kernel gradient are modified in order to restore consistency. For a given particle $a$, it reads 
\begin{align}
	\widetilde{\nabla W}_{ab} &= \bm{L}_a \nabla W_{ab},
\end{align}
\noindent where $\bm{L}_a = \left( \sum_{b\in\Lambda_{a}}\frac{m_a}{\rho_{a}}\nabla W_{ab} \otimes (\bm{x}_b - \bm{x}_a) \right)^{-1}$. Note that the tilde notation will be dropped in the rest of paper although the kernel gradient correction will be always used.

Second, the shifting technique for multiphase flows introduced in~\cite{Mokos2016} has also been used to maintain a good particles distribution. At the end of every timestep, all particles are shifted by a distance $\delta \bm{r}^s$ from their original position. This shifting distance of a particle $a$ is evaluated by doing
\begin{equation}
\delta \bm{r}^s_a =
\left\{ 
\begin{array}{l l}
 - D_a \nabla C_a\quad\text{if }a\in\text{light phase}\\
 - D_a \left(\frac{\partial C_a}{\partial s}\bm{s} + \alpha_n \left(\frac{\partial C_a}{\partial n}\bm{n} - \beta_n \right) \right)\quad\text{if }a\in\text{heavy phase}
\end{array} \right.
\end{equation}
\noindent where $C_a = \sum_{b\in\Lambda_{a}}\frac{m_a}{\rho_{a}}W_{ab}$ is the particle concentration, $\nabla C_a=\sum_{b\in\Lambda_{a}}\frac{m_a}{\rho_{a}} (C_b-C_a) \nabla W_{ab}$ is the particle concentration gradient, $D_a$ is the diffusion coefficient, $\bm{s}$ and $\bm{n}$ are respectively the tangent and normal vectors to the interface light/heavy phase (with $\bm{n}$ oriented towards the light phase), $\beta_n$ is a reference concentration gradient (taken equal to its initial value) and $\alpha_n$ is the normal diffusion parameter and is set equal to $0.1$.
The diffusion coefficient $D_a$ is computed as follows
\begin{equation}
D_a = A_s \lvert \bm{u}_a \rvert \Delta t
\end{equation}
\noindent where $A_s$ is a parameter set to $2$, $\bm{u}_a$ is the velocity of particle $a$, and $\Delta t$ is the timestep.

Third, as reported by several authors~\cite{colagrossi2003,grenier2009,szewc2013,ghai2017}, multiphase SPH can suffer from sub-kernel micro-mixing phenomena. Around the interface, within a distance corresponding ot the range of the kernel smoothing, particles have a tendency to mix. It is due to the fact that there is no mechanism ensuring phases immiscibility in the surface tension's continuum surface stress model. As suggested by the previously mentioned authors, we introduce a small repulsive force between phases as follows

\begin{equation}
\bm{F}^{corr}_a = \varepsilon \sum_{b\in\Lambda_{a},b\notin\Omega_{a}} \left(\frac{1}{\Theta_a^2} + \frac{1}{\Theta_b^2}\right) \nabla_a W_{ab},
\end{equation}
\noindent where $\varepsilon=\frac{L_{\text{ref}}}{h}$ for all simulations as suggested~ in~\cite{Szewc2016} where $L_{\text{ref}}$ is a reference length, typically the diameter for pipes. Note that the alternative formulation given by $\bm{F}^{corr}_a = \varepsilon \sum_{b\in\Lambda_{a},b\notin\Omega_{a}} \left(\frac{p_{a}}{\Theta_a^2} + \frac{p_{b}}{\Theta_b^2}\right) \nabla_a W_{ab}$ with $\varepsilon =0.1$ produces roughly the same force magnitude at the interface. The impact of this corrective force on the simulation of the same test cases studied in this paper can be found in~\cite{douillet2018}.

\subsection{Time Integration}

As with other numerical methods, any time integration scheme could be used. For the SPH method, both the Velocity Verlet and Predictor-Corrector Leapfrog schemes have proven popular. For this work, the Predictor-Corrector Leapfrog scheme was adopted as it was shown in~\cite{Tartakovsky2015} to be more stable than the Velocity Verlet scheme in the presence of boundary conditions.

The algorithm is described hereafter. For every particle $a$,

\begin{enumerate}
\item Predictor Step
\begin{equation}
\bm{u}^{n} = 
\left\{ 
\begin{array}{l l}
\bm{u}^{0}\quad\text{if }t=0\\
\bm{u}^{n-\frac{1}{2}} + \frac{\Delta t}{2} {\frac{D \bm{u}}{D t}}^{n-1}\quad\text{if }t>0
\end{array} \right.
\end{equation}
\item Compute $\rho^{n}_a$ and $p^{n}_a$ using the corresponding expressions in equation~\eqref{fullformulation}.
\item Evaluate $\frac{D \bm{u}}{D t}^{n}$ using the momentum equation in~\eqref{fullformulation}.
\item Corrector Step
\begin{equation}
\bm{u}^{n+\frac{1}{2}} =
\left\{ 
\begin{array}{l l}
\bm{u}^{n} + \frac{\Delta t}{2} {\frac{D \bm{u}}{D t}}^{n}\quad\text{if }t=0\\
\bm{u}^{n-\frac{1}{2}} + \Delta t {\frac{D \bm{u}}{D t}}^{n}\quad\text{if }t>0
\end{array} \right.
\end{equation}
\begin{equation}
\bm{x}^{n+1} = \bm{x}^{n} + \Delta t \bm{u}^{n+\frac{1}{2}}
\end{equation}
\end{enumerate}

The time step $\Delta t$ has to respect the Courant-Friedrichs-Lewy (CFL) criteria to ensure a stable evolution of the system e.g.

\begin{equation}
	\Delta t = \min \left( \Delta t_{\text{visc}}, \Delta t_{\text{grav}}, \Delta t_{\text{speed}}, \Delta t_{\text{st}} \right),
\end{equation}

\noindent where, following~\cite{morris1999surfacetension}, we have 

\begin{equation}
\left\{ 
\begin{array}{l l}
	\Delta t_{\text{visc}} &= 0.125\, \min_{\alpha \in \{1,\ldots,N_{\text{phases}}\}} \frac{ h^2 \rho^0_{\alpha}}{\mu_{\alpha}},\\
	\Delta t_{\text{grav}} &= 0.25\, \sqrt{\frac{h}{\lvert\bm{g} \rvert}},\\
	\Delta t_{\text{speed}} &= 0.25\, \min_{\alpha \in \{1,\ldots,N_{\text{phases}}\}} \frac{h}{c_{\alpha}},\\
	\Delta t_{\text{st}} &= 0.25\, \min_{\alpha,\beta \in \{1,\ldots,N_{\text{phases}}\}} \sqrt{ \frac{h^3 \rho^0_{\alpha} } {2 \pi \sigma^{\alpha \beta}} }.
\end{array} \right.
\end{equation}

A recent article~\cite{violeau2014timestep} investigated in detail what is the maximum admissible timestep in the WCSPH context. 

\section{Boundary Conditions}\label{sec:bc}

\begin{figure}[bthp]
	\begin{center}
		\resizebox{0.65\textwidth}{!}{\input{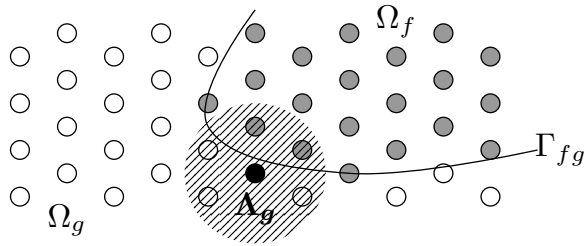}}
		\caption{Schematic of a ghost particle $g$ (in black) and its associated support domain $\Lambda_g$ (hatched area) intersecting with the fluid domain $\Omega_f$ (gray particles) and the ghost domain $\Omega_g$ (white particles) separated by $\Gamma_{fg}$.}
		\label{bc_dummy}
	\end{center}
\end{figure}

Because SPH is a collocation method, the treatment of boundary conditions is far from being trivial and is an active area of research within the community. The SPH European Research Interest Community~(SPHERIC) even made this topic one of its Grand Challenges for the development of the SPH method (\url{http://spheric-sph.org}).

Originally, in order to impose wall Boundary Conditions~(BC), repulsive forces based on the Lennard-Jones potential were used~\cite{monaghan2005}. However, it was shown in~\cite{Ferrand2012} that it induces strong spurious behaviors near the boundaries. Today, most models are imposing wall BC through the use of ghost particles or mirror particles~\cite{morris1997modeling,colagrossi2003,oger2006,Yildiz2009}. Recently, a new approach using semi-analytical~BC was introduced in~\cite{Ferrand2012, Leroy2014,ferrand2017} for both wall and inlet/outlet~BC. However, this semi-analytical approach is significantly more complex and time consuming, we preferred to use a simpler approach in view of the simple test cases that we consider in this work.

In this work, we used several types of~BC : no-slip wall (superscript $w$) and inlet/outlet (superscripts $in$ and $out$). For no-slip wall BC, the ghost particle method has been used along with the following prescribed values for the pressure $p^w$, density $\rho^w$ and velocity $\bm{v}^w$ for a given ghost particle $g$

\begin{align}
{p^w_{g}}&=\frac{1}{V_{ga}}\sum_{a\in\Omega_{f}\cap\Lambda_{g}}p_{a}\frac{m_a}{\rho_{a}}W_{ga},\\
{\rho^w_{g}}&=\frac{1}{V_{ga}}\sum_{a\in\Omega_{f}\cap\Lambda_{g}}\rho_{a}\frac{m_a}{\rho_{a}}W_{ga},\\
{\bm{v}^w_{g}}&=\frac{-1}{V_{ga}}\sum_{a\in\Omega_{f}\cap\Lambda_{g}}\bm{v}_{a}\frac{m_a}{\rho_{a}}W_{ga},
\end{align}

\noindent with $V_{ga}=\sum_{a\in\Omega_{f}\cap\Lambda_{g}}\frac{m_a}{\rho_{a}}W_{ga}$,  $\Omega_{f}$ the set of fluid particles and $\Lambda_{g}$ the set of neighboring particles of ghost particle $g$. A schematic drawn on Figure~\ref{bc_dummy} helps to visualize what is the intersection $\Omega_{f}\cap\Lambda_{g}$.

Inlet/outlet BC have also been subject to many investigations among SPH researchers. The main issue being that the naive way to implement those inlet/outlet BC results in spurious reflected waves~\cite{permeable,dong2014,khorasanizade2016,kunz2016,alvaro2017}. However, to the best of our knowledge, none of them addresses the issue of inlet/outlet BC for multiphase flows. In this paper, it has been decided to use the idea presented in~\cite{tafuni2018} and adapt it to multiphase SPH. To this end, the inlet and outlet boundaries are extended with a buffer layer of size $\kappa h$ to ensure a full kernel support. At the inlet, the goal is to inject the particles with a prescribed velocity profile. On the contrary, at the outlet, the particles need to leave the domain smoothly while imposing a prescribed pressure profile (or density since they are connected through equation~\eqref{tait_eos}).
On one hand, a particle $i$ in the inlet buffer is moving with a prescribed velocity profile $\bm{v}^p$ and it carries the following values of pressure $p^{in}$, density $\rho^{in}$ and velocity $\bm{v}^{in}$ 

\begin{align}
{p^{in}_{i}}&=\frac{1}{V_{ia}}\sum_{a\in\Omega_{f}\cap\Lambda_{i}}p_{a}\frac{m_a}{\rho_{a}}W_{ia},\\
{\rho^{in}_{i}}&=\frac{1}{V_{ia}}\sum_{a\in\Omega_{f}\cap\Lambda_{i}}\rho_{a}\frac{m_a}{\rho_{a}}W_{ia},\\
{\bm{v}^{in}_{i}}&= 2\bm{v}^p - \frac{1}{V_{ia}}\sum_{a\in\Omega_{f}\cap\Lambda_{i}}\bm{v}_{a}\frac{m_a}{\rho_{a}}W_{ia}.
\end{align}

\noindent with $V_{ia}=\sum_{a\in\Omega_{f}\cap\Lambda_{i}}\frac{m_a}{\rho_{a}}W_{ia}$ and $\Lambda_{i}$ the set of neighboring particles of inlet particle $i$.

On the other hand, at the outlet, a particle $o$ in the buffer is moved according to a smoothed convective velocity $\bm{v}^{conv}$. This idea is taken from~\cite{alvaro2017}. For example, if the outlet boundary is vertical and the flow leaves along the $x$ direction, it reads

\begin{equation}\label{conv}
 {\bm{v}^{out,conv}_{o}}=\frac{1}{V'_{oa}}\sum_{a\in\Lambda_{o}}\bm{v}_{a}\frac{m_a}{\rho_{a}}W_{oa},
\end{equation}

\noindent with $V'_{oa}=\sum_{a\in\Lambda_{o}}\frac{m_a}{\rho_{a}}W_{oa}$ the set of neighboring particles of outlet particle $o$. Note that in equation ~\eqref{conv}, the summation is over the full kernel support $\Lambda_{o}$ including fluid and outlet particles and not only over the intersection $\Omega_{f}\cap\Lambda_{o}$.
Besides, particle $o$ also carries the following values of pressure $p^{out}$, density $\rho^{out}$ and velocity $\bm{v}^{out}$
 
 \begin{align}
 {p^{out}_{o}}&= 2p^p - \frac{1}{V_{oa}}\sum_{a\in\Omega_{f}\cap\Lambda_{o}}p_{a}\frac{m_a}{\rho_{a}}W_{oa},\\
 {\rho^{out}_{o}}&= 2\rho^p - \frac{1}{V_{oa}}\sum_{a\in\Omega_{f}\cap\Lambda_{o}}\rho_{a}\frac{m_a}{\rho_{a}}W_{oa},\\
 {{v}^{out}_{o,x}}&=\frac{1}{V_{oa}}\sum_{a\in\Omega_{f}\cap\Lambda_{o}}{v}_{a,x}\frac{m_a}{\rho_{a}}W_{oa},\\
 {{v}^{out}_{o,y}}&=\frac{-1}{V_{oa}}\sum_{a\in\Omega_{f}\cap\Lambda_{o}}{v}_{a,y}\frac{m_a}{\rho_{a}}W_{oa},
 \end{align}

\noindent with $V_{oa}=\sum_{a\in\Omega_{f}\cap\Lambda_{o}}\frac{m_a}{\rho_{a}}W_{oa}$, $p^p$ and $\rho^p$ the prescribed pressure and density. Concerning the velocity, note that null cross velocities (here $v_y$) are enforced to ensure a divergence free velocity field at the outlet.

One last but important point is the treatment of the interface stress $\Pi_{a}$ introduced in equations~\eqref{st_pi}-\eqref{st_piab}. After experimenting with different approaches, we concluded that the best option is to explicitly calculate the interface stress even within the buffer areas (no extrapolation) to guarantee a clean interface, especially at the outlet where the interface position is not known a priori.

\begin{figure}[bthp]
	\centering
	\subfloat[\label{bc_validation_10ka}] {\includegraphics[width=0.333\textwidth,height=0.04\textwidth]{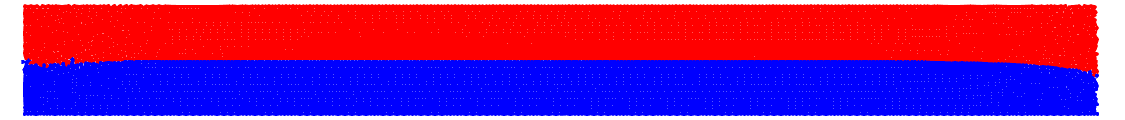} \includegraphics[width=0.333\textwidth,height=0.04\textwidth]{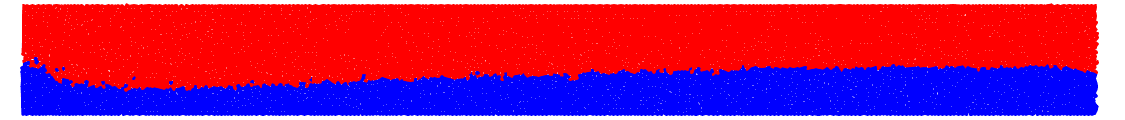} \includegraphics[width=0.333\textwidth,height=0.04\textwidth]{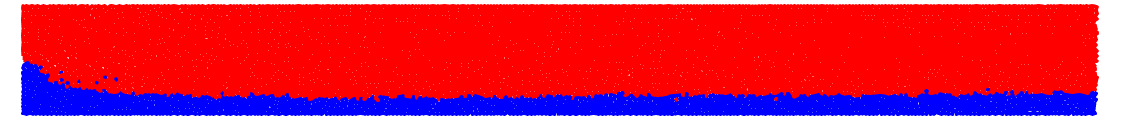}}\\	
	\subfloat[\label{bc_validation_10kb}] {\resizebox{0.333\textwidth}{!}{\input{IO_BC_Strat_10k_fig1}}}\hfill
	\subfloat[\label{bc_validation_10kc}] {\resizebox{0.333\textwidth}{!}{\input{IO_BC_Strat_10k_fig2}}}\hfill
	\subfloat[\label{bc_validation_10kd}] {\resizebox{0.333\textwidth}{!}{\input{IO_BC_Strat_10k_fig3}}}
	\caption {Case $L/\Delta r = 312$ ($10000$ particles). (a) Evolution of the phases distribution at $t=1~\second$, $t=10~\second$ and $t=30~\second$ (not at scale). (b) Evolution of the number of particles with time. (c) Evolution of the outlet pressure with time. (d) Evolution of the inlet velocities with time.}
	\label{bc_validation_10k}
\end{figure}

\begin{figure}[bthp]
	\centering
	\subfloat[\label{bc_validation_20ka}] {\includegraphics[width=0.333\textwidth,height=0.04\textwidth]{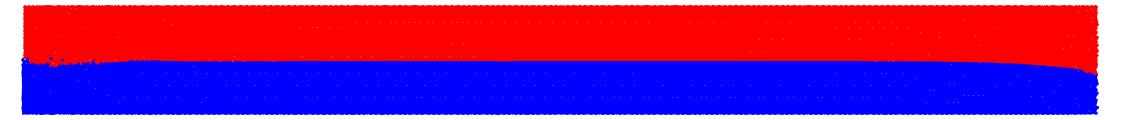} \includegraphics[width=0.333\textwidth,height=0.04\textwidth]{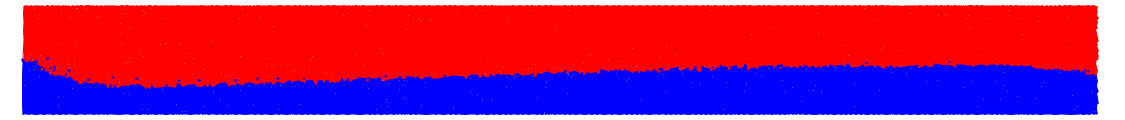} \includegraphics[width=0.333\textwidth,height=0.04\textwidth]{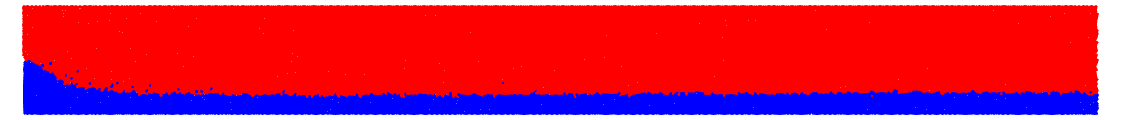}}\\
	\subfloat[\label{bc_validation_20kb}] {\resizebox{0.333\textwidth}{!}{\input{IO_BC_Strat_20k_fig1}}}\hfill
	\subfloat[\label{bc_validation_20kc}] {\resizebox{0.333\textwidth}{!}{\input{IO_BC_Strat_20k_fig2}}}\hfill
	\subfloat[\label{bc_validation_20kd}] {\resizebox{0.333\textwidth}{!}{\input{IO_BC_Strat_20k_fig3}}}
	\caption {Case $L/\Delta r = 444$ ($20000$ particles). (a) Evolution of the phases distribution at $t=1~\second$, $t=10~\second$ and $t=30~\second$  (not at scale). (b) Evolution of the number of particles with time. (c) Evolution of the outlet pressure with time. (d) Evolution of the inlet velocities with time.}
	\label{bc_validation_20k}
\end{figure}

\begin{figure}[bthp]
	\centering
	\subfloat[\label{bc_validation_50ka}] {\includegraphics[width=0.333\textwidth,height=0.04\textwidth]{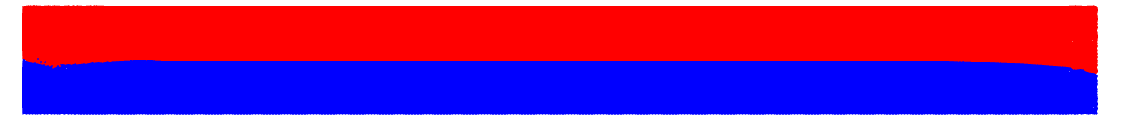} \includegraphics[width=0.333\textwidth,height=0.04\textwidth]{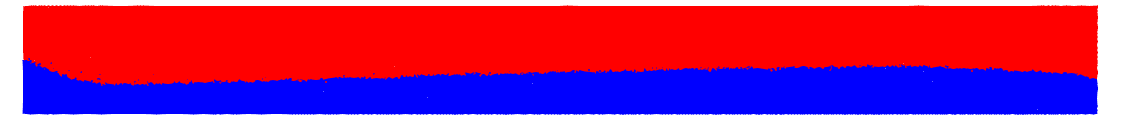} \includegraphics[width=0.333\textwidth,height=0.04\textwidth]{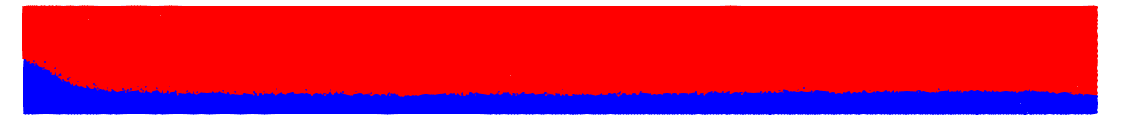}}\\
	\subfloat[\label{bc_validation_50kb}] {\resizebox{0.333\textwidth}{!}{\input{IO_BC_Strat_50k_fig1}}}\hfill
	\subfloat[\label{bc_validation_50kc}] {\resizebox{0.333\textwidth}{!}{\input{IO_BC_Strat_50k_fig2}}}\hfill
	\subfloat[\label{bc_validation_50kd}] {\resizebox{0.333\textwidth}{!}{\input{IO_BC_Strat_50k_fig3}}}
	\caption {Case $L/\Delta r = 704$ ($50000$ particles). (a) Evolution of the phases distribution at $t=1~\second$, $t=10~\second$ and $t=30~\second$  (not at scale). (b) Evolution of the number of particles with time. (c) Evolution of the outlet pressure with time. (d) Evolution of the inlet velocities with time.}
	\label{bc_validation_50k}
\end{figure}

In order to illustrate how this boundary condition implementation performs, a test case was simulated where a stratified flow ($50\%$ light upper phase and $50\%$ heavy lower phase) is injected with the following prescribed velocities : $u_g=1.4~\meter\per\second$ and $u_l=0.12~\meter\per\second$. At the outlet, the prescribed pressure is equal to the background pressure. The density ratio is $5$ and the viscosity ratio is $2$. Simulations are done with the following resolutions $L/\Delta r = 312$, $444$ and $704$ which corresponds approximately to $10000$, $20000$ and $50000$ particles. Several indicators are presented on Figures~\ref{bc_validation_10k} to~\ref{bc_validation_50k}. By 'Normalized Average Pressure/Velocity', we mean that the pressure/velocity is averaged among all particles over a distance of $\kappa h$ inside the fluid flow and over $10$ timesteps. Finally, it is divided by the prescribed value.

On Figures~\ref{bc_validation_10kb},~\ref{bc_validation_20kb} and~\ref{bc_validation_50kb}, we note that the number of particles within the fluid flow is maintained throughout the simulation. The prescribed velocity at the inlet is reasonably well reproduced with an error that stays within $\pm 5 \%$ for the light phase and $\pm 15\%$ for the heavy phase as shown on Figures~\ref{bc_validation_10kd},~\ref{bc_validation_20kd} and~\ref{bc_validation_50kd}. The difference between the two phases is likely due to the relatively low number of particles (especially in the $y$ direction) and to the case geometry where the heavy phase has to push against the light phase causing more disturbances in the velocity field for the heavy phase. On the other hand, at the outlet, the prescribed pressure is very well recovered with an maximum error of $\pm 1\%$ as presented on Figures~\ref{bc_validation_10kc},~\ref{bc_validation_20kc} and~\ref{bc_validation_50kc}. Besides, we observe that the errors are decreasing when the number of particles increases. These boundary conditions are not optimal but they perform reasonably well and are very easy to implement.
 
\section{Results}

There exists a large variety of flow regime maps to help characterize two-phase flow regime in pipes. For the particular case of horizontal pipes, one can mention Baker's map~\cite{Baker1953} and Barnea's map~\cite{Barnea1987}. In this work, Taitel and Dukler's map~\cite{Taitel1976} has been used to predict flow regimes.

\subsection{Validation for different flow regimes}\label{valid}
\begin{figure}[bthp]
	\begin{center}
		\resizebox{1.0\textwidth}{!}{\input{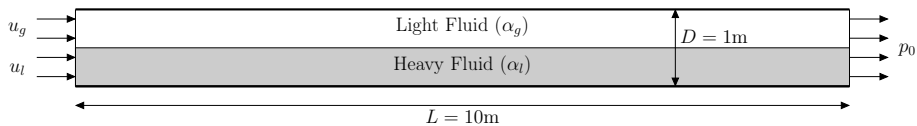}}
		\caption {The initial configuration for all cases (not at scale).}
		\label{init_geo}
	\end{center}
\end{figure}

\begin{figure}[bthp]
	\begin{center}
		\resizebox{0.75\textwidth}{!}{\input{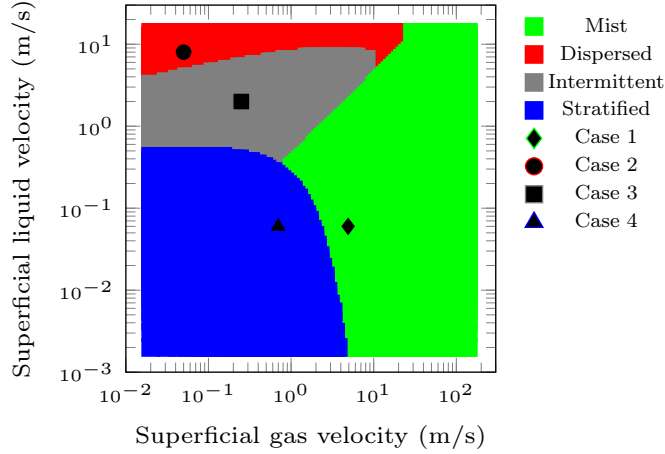}}
		\caption {Flow regime map adapted from~\cite{Taitel1976}}
		\label{flow_map}
	\end{center}
\end{figure}

\begin{figure}[bthp]
	\centering
	\subfloat[$L/\Delta r = 312$\label{flow_regime_if_corr_case1a}]{\includegraphics[width=0.333\textwidth,height=0.04\textwidth]{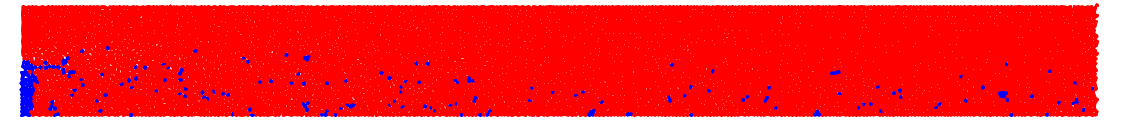}}\hfill
	\subfloat[$L/\Delta r = 444$\label{flow_regime_if_corr_case1b}]{\includegraphics[width=0.333\textwidth,height=0.04\textwidth]{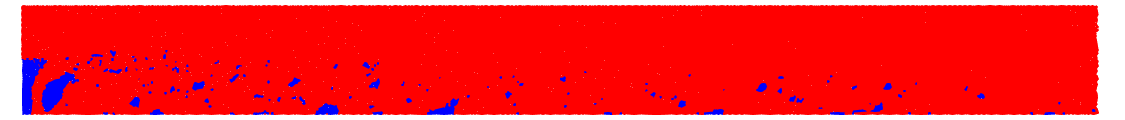}}\hfill
	\subfloat[$L/\Delta r = 704$\label{flow_regime_if_corr_case1c}]{\includegraphics[width=0.333\textwidth,height=0.04\textwidth]{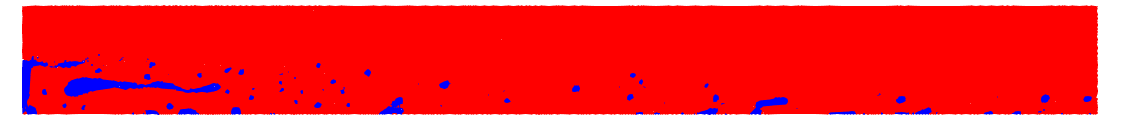}}\\
	\subfloat[$L/\Delta r = 312$\label{flow_regime_if_corr_case1d}]{\resizebox{0.333\textwidth}{!}{\input{Annular_10k_if_corr_fig1}}}\hfill
	\subfloat[$L/\Delta r = 444$\label{flow_regime_if_corr_case1e}]{\resizebox{0.333\textwidth}{!}{\input{Annular_20k_if_corr_fig1}}}\hfill
	\subfloat[$L/\Delta r = 704$\label{flow_regime_if_corr_case1f}]{\resizebox{0.333\textwidth}{!}{\input{Annular_50k_if_corr_fig1}}}\\
	\subfloat[$L/\Delta r = 312$\label{flow_regime_if_corr_case1g}]{\resizebox{0.333\textwidth}{!}{\input{Annular_10k_if_corr_fig2}}}\hfill
	\subfloat[$L/\Delta r = 444$\label{flow_regime_if_corr_case1h}]{\resizebox{0.333\textwidth}{!}{\input{Annular_20k_if_corr_fig2}}}\hfill
	\subfloat[$L/\Delta r = 704$\label{flow_regime_if_corr_case1i}]{\resizebox{0.333\textwidth}{!}{\input{Annular_50k_if_corr_fig2}}}\\
	\caption {Results for case 1 (Mist flow) : (a,b,c) Phases distribution at $t=30~\second$  (not at scale). (d,e,f) Evolution of the volume fractions at the outlet with time. (g,h,i) Evolution of the average pressure drop at the outlet with time.}
	\label{flow_regime_if_corr_case1}
\end{figure}

\begin{figure}[bthp]
	\centering
	\subfloat[$L/\Delta r = 312$\label{flow_regime_if_corr_case2a}]{\includegraphics[width=0.333\textwidth,height=0.04\textwidth]{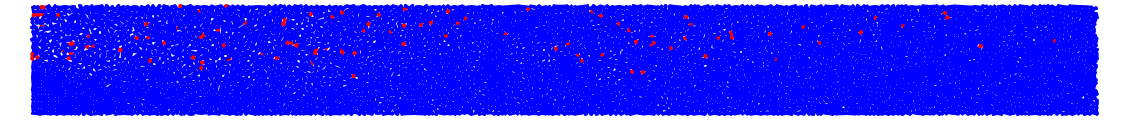}}\hfill
	\subfloat[$L/\Delta r = 444$\label{flow_regime_if_corr_case2b}]{\includegraphics[width=0.333\textwidth,height=0.04\textwidth]{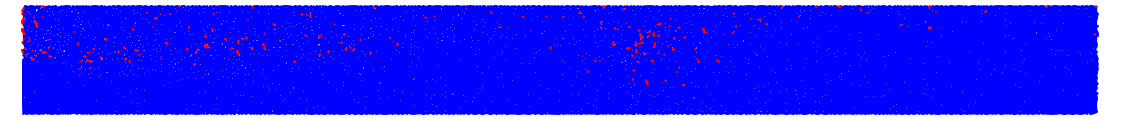}}\hfill
	\subfloat[$L/\Delta r = 704$\label{flow_regime_if_corr_case2c}]{\includegraphics[width=0.333\textwidth,height=0.04\textwidth]{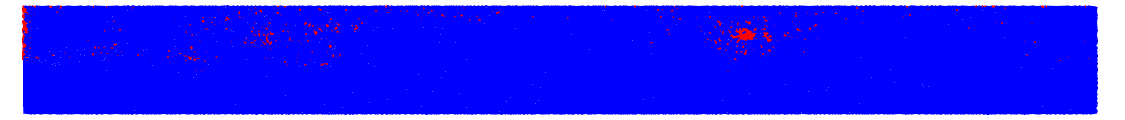}}\\
	\subfloat[$L/\Delta r = 312$\label{flow_regime_if_corr_case2d}]{\resizebox{0.333\textwidth}{!}{\input{Dispersed_10k_if_corr_fig1}}}\hfill
	\subfloat[$L/\Delta r = 444$\label{flow_regime_if_corr_case2e}]{\resizebox{0.333\textwidth}{!}{\input{Dispersed_20k_if_corr_fig1}}}\hfill
	\subfloat[$L/\Delta r = 704$\label{flow_regime_if_corr_case2f}]{\resizebox{0.333\textwidth}{!}{\input{Dispersed_50k_if_corr_fig1}}}\\
	\subfloat[$L/\Delta r = 312$\label{flow_regime_if_corr_case2g}]{\resizebox{0.333\textwidth}{!}{\input{Dispersed_10k_if_corr_fig2}}}\hfill
	\subfloat[$L/\Delta r = 444$\label{flow_regime_if_corr_case2h}]{\resizebox{0.333\textwidth}{!}{\input{Dispersed_20k_if_corr_fig2}}}\hfill
	\subfloat[$L/\Delta r = 704$\label{flow_regime_if_corr_case2i}]{\resizebox{0.333\textwidth}{!}{\input{Dispersed_50k_if_corr_fig2}}}\\
	\caption {Results for case 2 (Dispersed flow) : (a,b,c) Phases distribution at $t=30~\second$  (not at scale). (d,e,f) Evolution of the volume fractions at the outlet with time. (g,h,i) Evolution of the average pressure drop at the outlet with time.}
	\label{flow_regime_if_corr_case2}
\end{figure}

\begin{figure}[bthp]
	\centering
	\subfloat[$L/\Delta r = 312$\label{flow_regime_if_corr_case3a}]{\includegraphics[width=0.333\textwidth,height=0.04\textwidth]{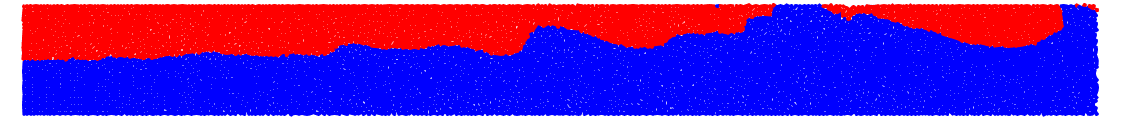}}\hfill
	\subfloat[$L/\Delta r = 444$\label{flow_regime_if_corr_case3b}]{\includegraphics[width=0.333\textwidth,height=0.04\textwidth]{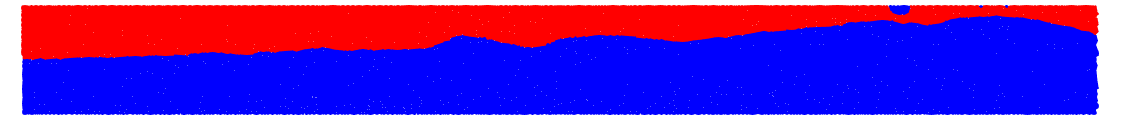}}\hfill
	\subfloat[$L/\Delta r = 704$\label{flow_regime_if_corr_case3c}]{\includegraphics[width=0.333\textwidth,height=0.04\textwidth]{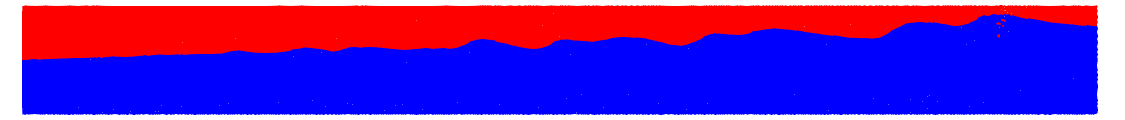}}\\
	\subfloat[$L/\Delta r = 312$\label{flow_regime_if_corr_case3d}]{\resizebox{0.333\textwidth}{!}{\input{Slug_10k_if_corr_fig1}}}\hfill
	\subfloat[$L/\Delta r = 444$\label{flow_regime_if_corr_case3e}]{\resizebox{0.333\textwidth}{!}{\input{Slug_20k_if_corr_fig1}}}\hfill
	\subfloat[$L/\Delta r = 704$\label{flow_regime_if_corr_case3f}]{\resizebox{0.333\textwidth}{!}{\input{Slug_50k_if_corr_fig1}}}\\
	\subfloat[$L/\Delta r = 312$\label{flow_regime_if_corr_case3g}]{\resizebox{0.333\textwidth}{!}{\input{Slug_10k_if_corr_fig2}}}\hfill
	\subfloat[$L/\Delta r = 444$\label{flow_regime_if_corr_case3h}]{\resizebox{0.333\textwidth}{!}{\input{Slug_20k_if_corr_fig2}}}\hfill
	\subfloat[$L/\Delta r = 704$\label{flow_regime_if_corr_case3i}]{\resizebox{0.333\textwidth}{!}{\input{Slug_50k_if_corr_fig2}}}\\
	\caption {Results for case 3 (Intermittent flow) : (a,b,c) Phases distribution at $t=30~\second$  (not at scale). (d,e,f) Evolution of the volume fractions at the outlet with time. (g,h,i) Evolution of the average pressure drop at the outlet with time.}
	\label{flow_regime_if_corr_case3}
\end{figure}

\begin{figure}[bthp]
	\centering
	\subfloat[$t=10~\second$\label{long_pipeb}]{\includegraphics[width=0.5\textwidth,height=0.04\textwidth]{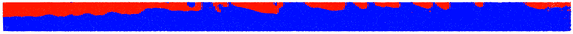}}\\
	\subfloat[$t=20~\second$\label{long_pipec}]{\includegraphics[width=0.5\textwidth,height=0.04\textwidth]{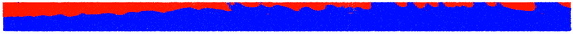}}\\
	\subfloat[$t=30~\second$\label{long_piped}]{\includegraphics[width=0.5\textwidth,height=0.04\textwidth]{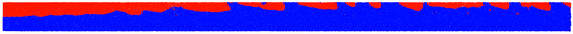}}\\
	\caption {Phases distribution at selected instants for case 3 (Intermittent flow) with a pipe of $20~\meter$ ($L/\Delta r = 312$, not at scale).}
	\label{long_pipe}
\end{figure}

\begin{figure}[bthp]
	\centering
	\subfloat[$L/\Delta r = 312$\label{flow_regime_if_corr_case4a}]{\includegraphics[width=0.333\textwidth,height=0.04\textwidth]{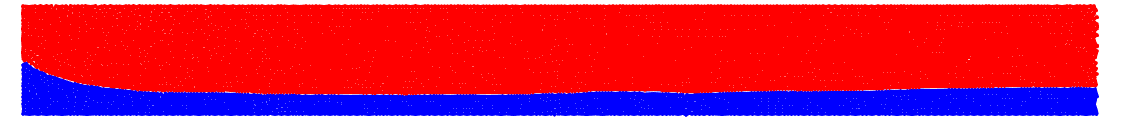}}\hfill
	\subfloat[$L/\Delta r = 444$\label{flow_regime_if_corr_case4b}]{\includegraphics[width=0.333\textwidth,height=0.04\textwidth]{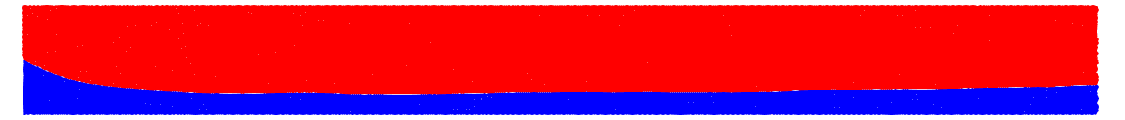}}\hfill
	\subfloat[$L/\Delta r = 704$\label{flow_regime_if_corr_case4c}]{\includegraphics[width=0.333\textwidth,height=0.04\textwidth]{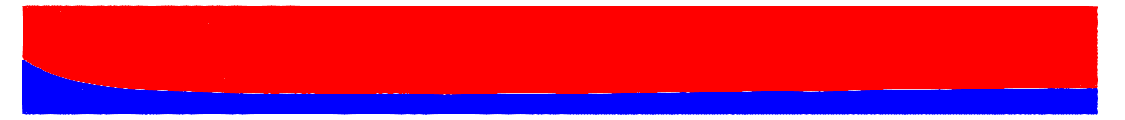}}\\
	\subfloat[$L/\Delta r = 312$\label{flow_regime_if_corr_case4d}]{\resizebox{0.333\textwidth}{!}{\input{Stratified_10k_if_corr_fig1}}}\hfill
	\subfloat[$L/\Delta r = 444$\label{flow_regime_if_corr_case4e}]{\resizebox{0.333\textwidth}{!}{\input{Stratified_20k_if_corr_fig1}}}\hfill
	\subfloat[$L/\Delta r = 704$\label{flow_regime_if_corr_case4f}]{\resizebox{0.333\textwidth}{!}{\input{Stratified_50k_if_corr_fig1}}}\\
	\subfloat[$L/\Delta r = 312$\label{flow_regime_if_corr_case4g}]{\resizebox{0.333\textwidth}{!}{\input{Stratified_10k_if_corr_fig2}}}\hfill
	\subfloat[$L/\Delta r = 444$\label{flow_regime_if_corr_case4h}]{\resizebox{0.333\textwidth}{!}{\input{Stratified_20k_if_corr_fig2}}}\hfill
	\subfloat[$L/\Delta r = 704$\label{flow_regime_if_corr_case4i}]{\resizebox{0.333\textwidth}{!}{\input{Stratified_50k_if_corr_fig2}}}\\
	\caption {Results for case 4 (Stratified flow) : (a,b,c) Phases distribution at $t=30~\second$  (not at scale). (d,e,f) Evolution of the volume fractions at the outlet with time. (g,h,i) Evolution of the average pressure drop at the outlet with time.}
	\label{flow_regime_if_corr_case4}
\end{figure}

\begin{figure}[bthp]
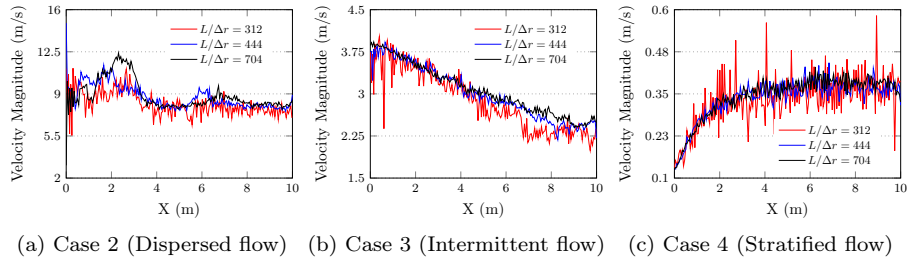

	\centering
	\subfloat[Case 2 (Dispersed flow)\label{velmagheavydisp}]{\resizebox{0.33\textwidth}{!}{\input{mesh_conv_velheavy_disp_old}}}
	\subfloat[Case 3 (Intermittent flow)\label{velmagheavyslug}]{\resizebox{0.33\textwidth}{!}{\input{mesh_conv_velheavy_slug_old}}}
	\subfloat[Case 4 (Stratified flow)\label{velmagheavystrat}]{\resizebox{0.33\textwidth}{!}{\input{mesh_conv_velheavy_strat_old}}}
	\caption {Heavy phase velocity magnitude along the pipe length at $t=30~\second$ in function of the particle resolution.}
	\label{velmagheavy}
\end{figure}

\begin{figure}[bthp]
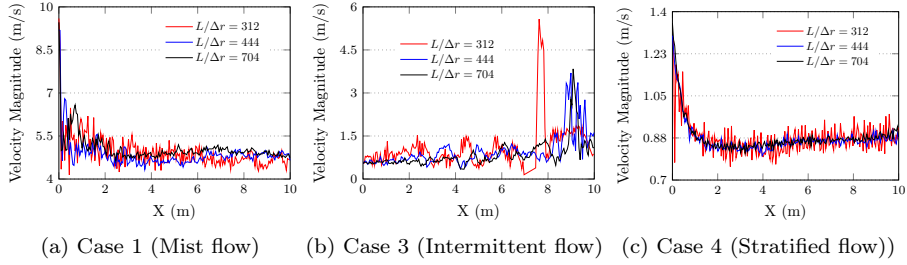

	\centering
	\subfloat[Case 1 (Mist flow)\label{velmaglightann}]{\resizebox{0.33\textwidth}{!}{\input{mesh_conv_vellight_ann_old}}}
	\subfloat[Case 3 (Intermittent flow)\label{velmaglightslug}]{\resizebox{0.33\textwidth}{!}{\input{mesh_conv_vellight_slug_old}}}
	\subfloat[Case 4 (Stratified flow))\label{velmaglightstrat}]{\resizebox{0.33\textwidth}{!}{\input{mesh_conv_vellight_strat_old}}}
	\caption {Light phase velocity magnitude along the pipe length at $t=30~\second$ in function of the particle resolution.}
	\label{velmaglight}
\end{figure}

We consider an horizontal pipe of diameter $D=1\meter$ and length $L=10D$. The light phase and heavy phase are denoted with a $g$ and $l$ subscript respectively. The flow enters from the inlet (left) and is assumed to be stratified with equal volume fraction for each phase $\alpha_g=\alpha_l=0.5$.  All the physical properties are summarized in Table~\ref{tab:mat_prop}. Using these properties, it is possible to plot the flow regime map, see Figure~\ref{flow_map}\footnote{In order to plot the map, one has to compute the Lockhart-Martelli~\cite{lockhart1949} parameter which depends on $n$, $m$, $C_g$ and $C_l$. In this study, we used $n=m=2$ and $C_g=C_l=0.042$}, and to pick four cases, one in each region, to be simulated. Those cases and their corresponding parameters are presented in Table~\ref{tab:case_prop} and marked by different symbols on Figure~\ref{flow_map}. All cases were simulated in 2D with three different resolutions $L/\Delta r = 312$, $444$ and $704$ which corresponds approximately to $10000$, $20000$ and $50000$ particles. The simulation time was $30~\second$. At the inlet, each phase is injected with a constant velocity corresponding to its superficial velocity $u_{g,l}^s = \alpha_{g,l} u_{g,l}$. At the outlet, a constant pressure equal to the background pressure is prescribed. The initial setup is presented on Figure~\ref{init_geo}. The final two-flow patterns for the different cases, the evolution of the volume fraction with time at the outlet and the average pressure drop between both ends of the pipe is presented on Figures~\ref{flow_regime_if_corr_case1} to \ref{flow_regime_if_corr_case4}. In addition, the evolution of the velocity magnitude along the pipe's length for each phase for the three particle resolutions considered is shown on Figures~\ref{velmagheavy} and \ref{velmaglight}.

\begin{table}[bthp]
	\centering
	\begin{adjustbox}{max width=\textwidth}
	\begin{tabular}{llll}
		\toprule
		\textbf{Property}         & \textbf{Light Phase}   & \textbf{Heavy Phase} & \textbf{Units}\\
		\midrule
		\textbf{Density} ($\rho$) & $1$         & $5$          & $\kilogram\per\meter^{3}$\\
		\textbf{Viscosity} ($\mu$)& $5\times 10^{-3}$ & $1\times 10^{-2}$  & $\pascal .\second$\\
		\textbf{Adiabatic index} ($\gamma$) & $7$ & $7$  & -\\
		\textbf{Surface Tension} ($\sigma^{nw}$) & \multicolumn{2}{l}{0.001} & $\newton\per\meter$\\
		\textbf{Contact Angle} ($\theta_c$) & \multicolumn{2}{l}{90} & $\degree$\\
		\textbf{Gravity} ($g_z$) & \multicolumn{2}{l}{-1} & $\meter\per\second^2$\\
		\bottomrule
	\end{tabular}
\end{adjustbox}
	\caption{Physical Properties}
	\label{tab:mat_prop}
\end{table}

\begin{table}[bthp]
	\begin{adjustbox}{max width=\textwidth}
	\begin{tabular}{llllll}
		\toprule
		\textbf{Property} & \textbf{Case 1}   & \textbf{Case 2} & \textbf{Case 3} & \textbf{Case 4} & \textbf{Units}\\
		\midrule
		\textbf{Flow pattern} & Mist & Dispersed & Intermittent & Stratified & -\\
		\textbf{Superficial velocities} ($u_g^s$,$u_l^s$) & ($4.9$,$0.06$) & ($0.05$,$8$) & ($0.25$,$2$) & ($0.7$,$0.06$) & $\meter\per\second$\\
		\textbf{Reynolds number} ($Re=\frac{(u_g^s+u_l^s)D}{\nu_g \alpha_g + \nu_l \alpha_l}$) & $1417.14$ & $2300$ & $642.86$ & $217.14$ & -\\
		\textbf{Friedel Pressure Drop Prediction} ($\Delta p^{\text{Friedel}}$) & $20.931$ & $98.513$ & $11.961$ & $4.1$ & $\pascal$\\
		\bottomrule
	\end{tabular}
\end{adjustbox}
	\caption{Cases properties}
	\label{tab:case_prop}
\end{table}

For information purposes, Friedel correlation~\cite{friedel1979} has been used to evaluate the expected pressure drops for the four considered cases of Figure~\ref{flow_map}. These values reported in Table~\ref{tab:case_prop} are to be taken with caution as the Friedel correlation has proven not to be very reliable for rectangular channels, separated flows and/or viscous fluids~\cite{MllerSteinhagen1986,wambsganss1992two,Spedding2006}. Besides, on the same table, we also report the average Reynolds number for the considered geometry and flow conditions. Note that this is an average value. Depending on how it is computed, the Reynolds number could be locally much higher than the given value.

On Figures~\ref{flow_regime_if_corr_case1a} to~\ref{flow_regime_if_corr_case1c}, one can observe that a mist flow pattern where the heavy phase is scattered by the light phase. It is confirmed by Figures~\ref{flow_regime_if_corr_case1d} to~\ref{flow_regime_if_corr_case1f} where we see that the light phase outlet volume fraction goes to $1$ after a transient period. Note that, in 3D, this mist pattern could be an annular flow under certain conditions. As the number of particles increases, Figures~\ref{flow_regime_if_corr_case1g} to~\ref{flow_regime_if_corr_case1i} show that the pressure drop level decreases and appears to stabilize around $\approx 100 ~\pascal$. This is higher than predicted by Friedel correlation. Note that turbulence is not included in the present model whereas the Reynolds number $Re$ is becoming turbulent and that the turbulent viscosity would contribute to stabilize the pressure field. For turbulence modeling in an SPH context, one can refer to~\cite{shao2005,violeau2006,rogers2008} or~\cite{monaghan2011}.

On Figures~\ref{flow_regime_if_corr_case2a} to~\ref{flow_regime_if_corr_case2c}, pictures show a typical dispersed bubbly flow pattern where the light phase is spread-out by the heavy phase. Figures~\ref{flow_regime_if_corr_case2d} to~\ref{flow_regime_if_corr_case2f} support this claim as one can note that the heavy phase outlet volume fraction quickly goes to $1$ after a transient period. On Figures~\ref{flow_regime_if_corr_case2g} to~\ref{flow_regime_if_corr_case2i}, the pressure drop evolution with time presents large oscillations and its average level varies strongly with the particles' resolution (from $\approx 500 ~\pascal$ to $\approx 0 ~\pascal$). It even becomes negative at some instants testifying the occurrence of important recirculation areas near the light phase inlet. It is also higher than what Friedel correlation predicts. However, as stated before, turbulence effects are not taken into account whereas the Reynolds number $Re$ is typically turbulent. We believe the turbulent viscosity would help to stabilize the pressure field.

On Figures~\ref{flow_regime_if_corr_case3a} to~\ref{flow_regime_if_corr_case3c}, we can see that an intermittent flow is established as expected.  The intermittent character of the flow pattern is later confirmed by the volume fraction time series of Figures~\ref{flow_regime_if_corr_case3d} to~\ref{flow_regime_if_corr_case3f} where we can see that the light phase volume fraction is strongly oscillating between $0.25$ and $0$. It means that long bubbles are generated at a given frequency which corresponds to the definition of a slug flow. On Figures~\ref{flow_regime_if_corr_case3g} to~\ref{flow_regime_if_corr_case3i}, the pressure drop average level decreases when the number of particles increases, varying from $\approx 30 ~\pascal$ to $\approx 5 ~\pascal$. Friedel correlation predicts a pressure drop of $\approx 10 ~\pascal$ which is of the same order of magnitude. It goes under $0 ~\pascal$ during brief instants or during the transient phase because of the recirculation areas near the inlet. As shown on Figure~\ref{long_pipe}, the intermittent flow presented in this plots would evolve towards a fully developed slug flow if the pipe was longer.

Finally, on Figures~\ref{flow_regime_if_corr_case4a} to~\ref{flow_regime_if_corr_case4c}, we observe a fully developed stratified flow. Figures~\ref{flow_regime_if_corr_case4d} to~\ref{flow_regime_if_corr_case4f} show that the phase distribution is adjusting itself with time to reach a periodic steady state where the light phase volume fraction is $\approx 0.25$ and the heavy phase volume fraction is $\approx 0.75$. The pressure drop evolution presented on Figures~\ref{flow_regime_if_corr_case4g} to~\ref{flow_regime_if_corr_case4i} show much smaller oscillations than the other cases because the flow is evolving at a smaller speed. Its average level goes from $\approx 7.5 ~\pascal$ for the lowest resolution to $\approx 2.5 ~\pascal$ for the highest resolution. Friedel correlation gives an expected pressure drop of $\approx 4 ~\pascal$ which is of the same order of magnitude.

To sum up, one can observe that our current implementation of SPH is able to reproduce the two-phase flow patterns predicted by the flow map of Figure~\ref{flow_map}. Moreover, increasing the number of particles helps to reduce the pressure field oscillations while reproducing the same physics.

Moreover, on Figures~\ref{velmagheavy} and \ref{velmaglight}, we provide plots showing the evolution of the velocity magnitude along the pipe length at $t=30~\second$ to verify that we do have a convergence of the velocity field when the number of particles increases. To obtain these plots, the velocity magnitude has been averaged along the pipe's height for each phase. Note that, for the mist case (respectively the dispersed case), we do not show the heavy phase velocity (respectively the light phase velocity) since there are not enough particles of that phase to be considered within the pipe's height. From a general point of view, what we observed is that, as the resolution increases, the velocity field tends to become more stable solution, presenting smaller oscillations and converging towards a steady state. This is particularly clear for the stratified case of Figures~\ref{velmagheavystrat} and \ref{velmaglightstrat}. For the intermittent case, the velocity field also depends on the distribution of the bubbles of light phase. For instance, the peak on Figure~\ref{velmagheavyslug} for the lowest resolution corresponds to the tail of the bubble of Figure~\ref{flow_regime_if_corr_case3a}. This bubble just got formed and is self-adjusting its shape under the effect of surface tension, hence the peak in velocity. Concerning the mist and dispersed case of Figures~\ref{velmaglightann} and \ref{velmagheavydisp}, we do not recover the prescribed velocity of the dominant phase at the inlet. It is because we average the velocity along the pipe's height so that we include the recirculation areas at the entry which tends to reduce the magnitude of the velocity field. It is difficult to see a convergence of the pressure drops' evolution when refining the resolution in Figures~\ref{flow_regime_if_corr_case1} to \ref{flow_regime_if_corr_case4} because of the inherent pressure noise due to the weakly compressible SPH and more importantly because of the pressure waves reflections at the boundaries. However, we can visually observe that convergence in the velocity field especially for the steadier cases such as the stratified case.


\subsection{Transitions}\label{trans}

\begin{figure}[bthp]
	\begin{center}
		\resizebox{0.75\textwidth}{!}{\input{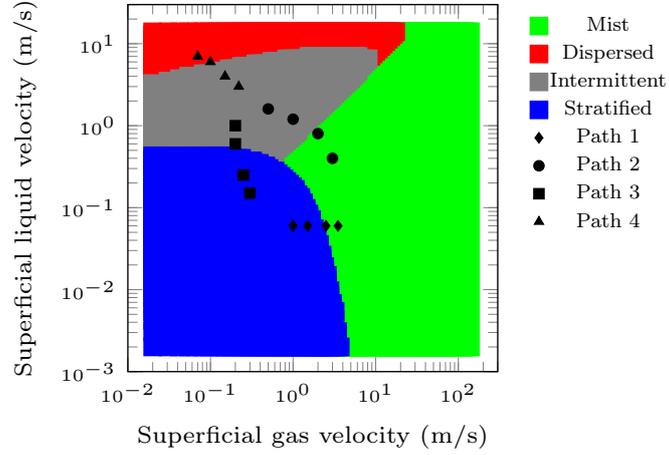}}
		\caption {Four different paths from one flow pattern to another.}
		\label{flow_map2}
	\end{center}
\end{figure}

\begin{figure}[bthp]
	\centering
	\subfloat[$\#1$\label{flow_transitions_if_corr_path1a}]{\includegraphics[width=0.25\textwidth,height=0.04\textwidth]{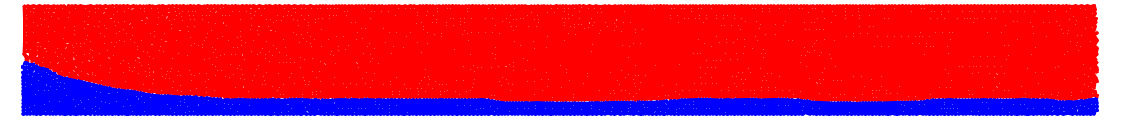}}\hfill
	\subfloat[$\#2$\label{flow_transitions_if_corr_path1b}]{\includegraphics[width=0.25\textwidth,height=0.04\textwidth]{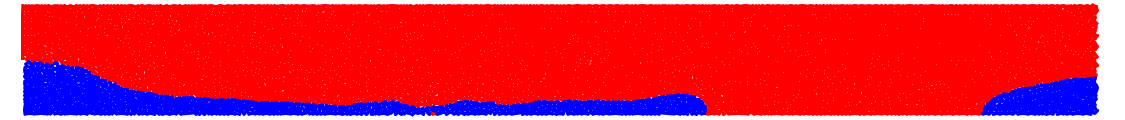}}\hfill
	\subfloat[$\#3$\label{flow_transitions_if_corr_path1c}]{\includegraphics[width=0.25\textwidth,height=0.04\textwidth]{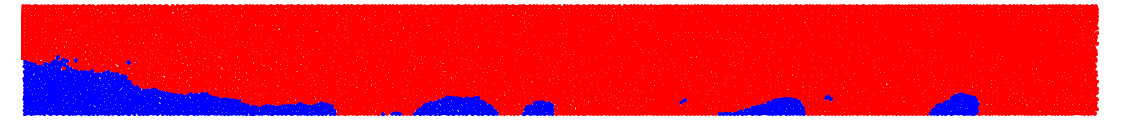}}\hfill
	\subfloat[$\#4$\label{flow_transitions_if_corr_path1d}]{\includegraphics[width=0.25\textwidth,height=0.04\textwidth]{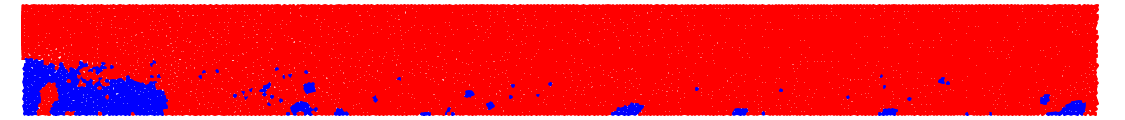}}\\
	\subfloat[$\#1$\label{flow_transitions_if_corr_path1e}]{\resizebox{0.25\textwidth}{!}{\input{strat2ann1_if_corr_fig1}}}
	\subfloat[$\#2$\label{flow_transitions_if_corr_path1f}]{\resizebox{0.25\textwidth}{!}{\input{strat2ann2_if_corr_fig1}}}
	\subfloat[$\#3$\label{flow_transitions_if_corr_path1g}]{\resizebox{0.25\textwidth}{!}{\input{strat2ann3_if_corr_fig1}}}
	\subfloat[$\#4$\label{flow_transitions_if_corr_path1h}]{\resizebox{0.25\textwidth}{!}{\input{strat2ann4_if_corr_fig1}}}\\
	\subfloat[$\#1$\label{flow_transitions_if_corr_path1i}]{\resizebox{0.25\textwidth}{!}{\input{strat2ann1_if_corr_fig2}}}
	\subfloat[$\#2$\label{flow_transitions_if_corr_path1j}]{\resizebox{0.25\textwidth}{!}{\input{strat2ann2_if_corr_fig2}}}
	\subfloat[$\#3$\label{flow_transitions_if_corr_path1k}]{\resizebox{0.25\textwidth}{!}{\input{strat2ann3_if_corr_fig2}}}
	\subfloat[$\#4$\label{flow_transitions_if_corr_path1l}]{\resizebox{0.25\textwidth}{!}{\input{strat2ann4_if_corr_fig2}}}\\
	\caption {Results for path 1 (Stratified to Mist) : (a,b,c,d) Phases distribution at $t=30~\second$  (not at scale). (e,f,g,h) Evolution of the volume fractions at the outlet with time. (i,j,k,l) Evolution of the average pressure drop at the outlet with time.}
	\label{flow_transitions_if_corr_path1}
\end{figure}

\begin{figure}[bthp]
	\centering
	\subfloat[$\#1$\label{flow_transitions_if_corr_path2a}]{\includegraphics[width=0.25\textwidth,height=0.04\textwidth]{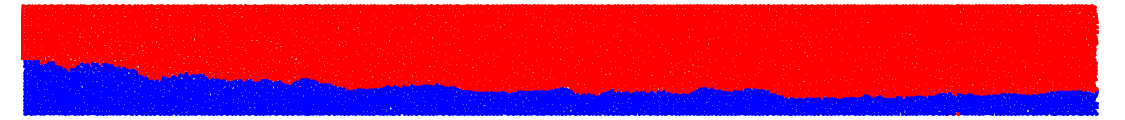}}\hfill
	\subfloat[$\#2$\label{flow_transitions_if_corr_path2b}]{\includegraphics[width=0.25\textwidth,height=0.04\textwidth]{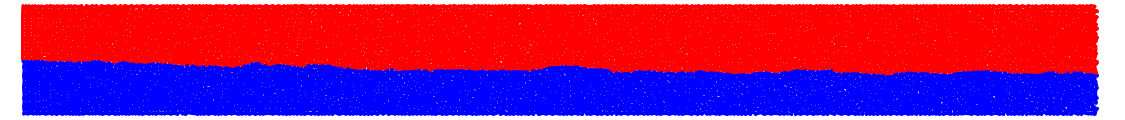}}\hfill
	\subfloat[$\#3$\label{flow_transitions_if_corr_path2c}]{\includegraphics[width=0.25\textwidth,height=0.04\textwidth]{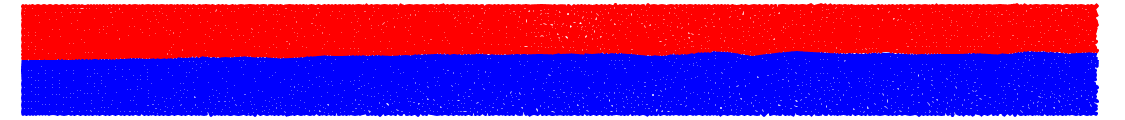}}\hfill
	\subfloat[$\#4$\label{flow_transitions_if_corr_path2d}]{\includegraphics[width=0.25\textwidth,height=0.04\textwidth]{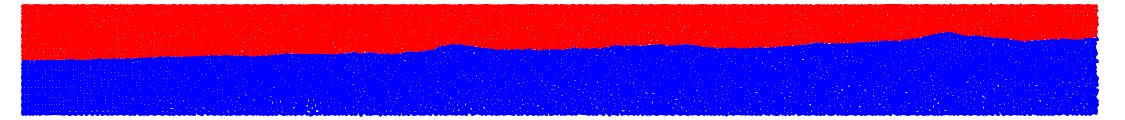}}\\
	\subfloat[$\#1$\label{flow_transitions_if_corr_path2e}]{\resizebox{0.25\textwidth}{!}{\input{ann2slug1_if_corr_fig1}}}
	\subfloat[$\#2$\label{flow_transitions_if_corr_path2f}]{\resizebox{0.25\textwidth}{!}{\input{ann2slug2_if_corr_fig1}}}
	\subfloat[$\#3$\label{flow_transitions_if_corr_path2g}]{\resizebox{0.25\textwidth}{!}{\input{ann2slug3_if_corr_fig1}}}
	\subfloat[$\#4$\label{flow_transitions_if_corr_path2h}]{\resizebox{0.25\textwidth}{!}{\input{ann2slug4_if_corr_fig1}}}\\
	\subfloat[$\#1$\label{flow_transitions_if_corr_path2i}]{\resizebox{0.25\textwidth}{!}{\input{ann2slug1_if_corr_fig2}}}
	\subfloat[$\#2$\label{flow_transitions_if_corr_path2j}]{\resizebox{0.25\textwidth}{!}{\input{ann2slug2_if_corr_fig2}}}
	\subfloat[$\#3$\label{flow_transitions_if_corr_path2k}]{\resizebox{0.25\textwidth}{!}{\input{ann2slug3_if_corr_fig2}}}
	\subfloat[$\#4$\label{flow_transitions_if_corr_path2l}]{\resizebox{0.25\textwidth}{!}{\input{ann2slug4_if_corr_fig2}}}\\
	\caption {Results for path 2 (Mist to Intermittent) : (a,b,c,d) Phases distribution at $t=30~\second$  (not at scale). (e,f,g,h) Evolution of the volume fractions at the outlet with time. (i,j,k,l) Evolution of the average pressure drop at the outlet with time.}
	\label{flow_transitions_if_corr_path2}
\end{figure}

\begin{figure}[bthp]
	\centering
	\subfloat[$\#1$\label{flow_transitions_if_corr_path3a}]{\includegraphics[width=0.25\textwidth,height=0.04\textwidth]{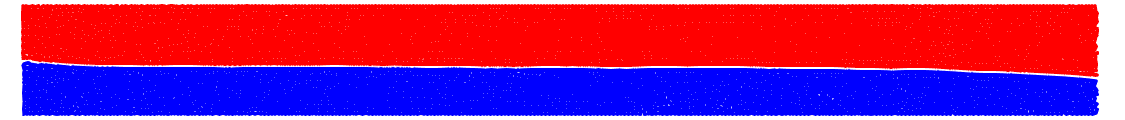}}\hfill
	\subfloat[$\#2$\label{flow_transitions_if_corr_path3b}]{\includegraphics[width=0.25\textwidth,height=0.04\textwidth]{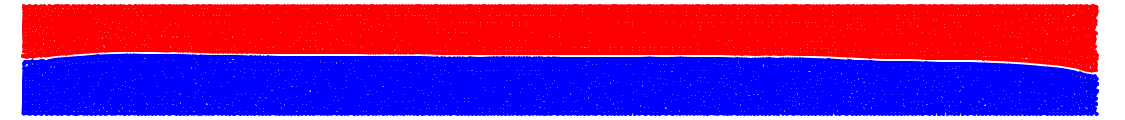}}\hfill
	\subfloat[$\#3$\label{flow_transitions_if_corr_path3c}]{\includegraphics[width=0.25\textwidth,height=0.04\textwidth]{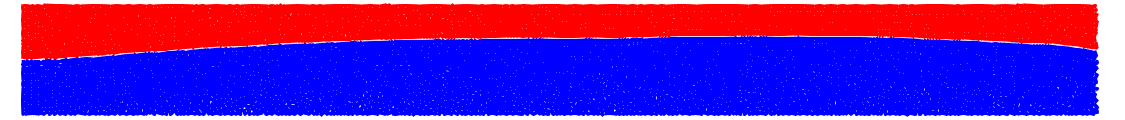}}\hfill
	\subfloat[$\#4$\label{flow_transitions_if_corr_path3d}]{\includegraphics[width=0.25\textwidth,height=0.04\textwidth]{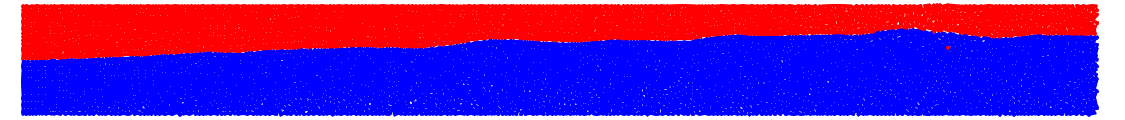}}\\
	\subfloat[$\#1$\label{flow_transitions_if_corr_path3e}]{\resizebox{0.25\textwidth}{!}{\input{strat2slug1_if_corr_fig1}}}
	\subfloat[$\#2$\label{flow_transitions_if_corr_path3f}]{\resizebox{0.25\textwidth}{!}{\input{strat2slug2_if_corr_fig1}}}
	\subfloat[$\#3$\label{flow_transitions_if_corr_path3g}]{\resizebox{0.25\textwidth}{!}{\input{strat2slug3_if_corr_fig1}}}
	\subfloat[$\#4$\label{flow_transitions_if_corr_path3h}]{\resizebox{0.25\textwidth}{!}{\input{strat2slug4_if_corr_fig1}}}\\
	\subfloat[$\#1$\label{flow_transitions_if_corr_path3i}]{\resizebox{0.25\textwidth}{!}{\input{strat2slug1_if_corr_fig2}}}
	\subfloat[$\#2$\label{flow_transitions_if_corr_path3j}]{\resizebox{0.25\textwidth}{!}{\input{strat2slug2_if_corr_fig2}}}
	\subfloat[$\#3$\label{flow_transitions_if_corr_path3k}]{\resizebox{0.25\textwidth}{!}{\input{strat2slug3_if_corr_fig2}}}
	\subfloat[$\#4$\label{flow_transitions_if_corr_path3l}]{\resizebox{0.25\textwidth}{!}{\input{strat2slug4_if_corr_fig2}}}\\
	\caption {Results for path 3 (Stratified to Intermittent) : (a,b,c,d) Phases distribution at $t=30~\second$  (not at scale). (e,f,g,h) Evolution of the volume fractions at the outlet with time. (i,j,k,l) Evolution of the average pressure drop at the outlet with time.}
	\label{flow_transitions_if_corr_path3}
\end{figure}

\begin{figure}[bthp]
	\centering
	\subfloat[$\#1$\label{flow_transitions_if_corr_path4a}]{\includegraphics[width=0.25\textwidth,height=0.04\textwidth]{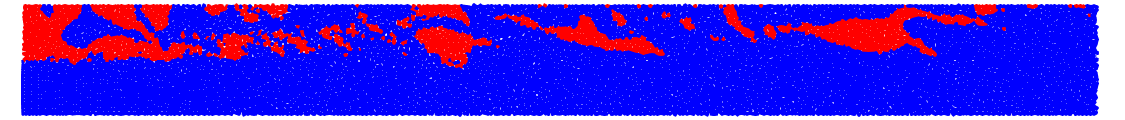}}\hfill
	\subfloat[$\#2$\label{flow_transitions_if_corr_path4b}]{\includegraphics[width=0.25\textwidth,height=0.04\textwidth]{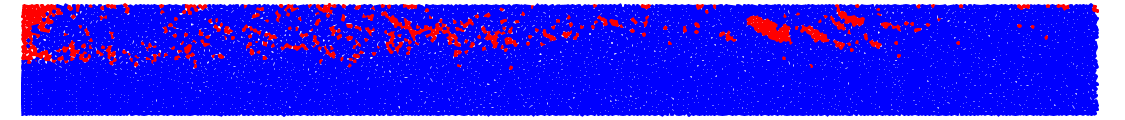}}\hfill
	\subfloat[$\#3$\label{flow_transitions_if_corr_path4c}]{\includegraphics[width=0.25\textwidth,height=0.04\textwidth]{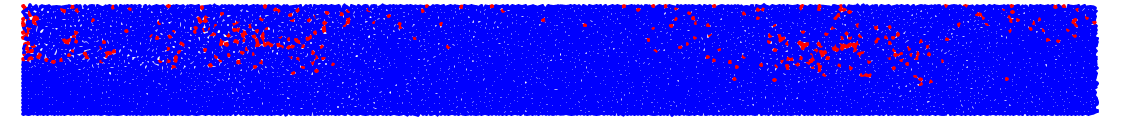}}\hfill
	\subfloat[$\#4$\label{flow_transitions_if_corr_path4d}]{\includegraphics[width=0.25\textwidth,height=0.04\textwidth]{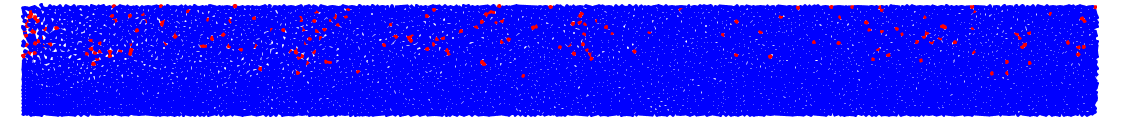}}\\
	\subfloat[$\#1$\label{flow_transitions_if_corr_path4e}]{\resizebox{0.25\textwidth}{!}{\input{slug2disp1_if_corr_fig1}}}
	\subfloat[$\#2$\label{flow_transitions_if_corr_path4f}]{\resizebox{0.25\textwidth}{!}{\input{slug2disp2_if_corr_fig1}}}
	\subfloat[$\#3$\label{flow_transitions_if_corr_path4g}]{\resizebox{0.25\textwidth}{!}{\input{slug2disp3_if_corr_fig1}}}
	\subfloat[$\#4$\label{flow_transitions_if_corr_path4h}]{\resizebox{0.25\textwidth}{!}{\input{slug2disp4_if_corr_fig1}}}\\
	\subfloat[$\#1$\label{flow_transitions_if_corr_path4i}]{\resizebox{0.25\textwidth}{!}{\input{slug2disp1_if_corr_fig2}}}
	\subfloat[$\#2$\label{flow_transitions_if_corr_path4j}]{\resizebox{0.25\textwidth}{!}{\input{slug2disp2_if_corr_fig2}}}
	\subfloat[$\#3$\label{flow_transitions_if_corr_path4k}]{\resizebox{0.25\textwidth}{!}{\input{slug2disp3_if_corr_fig2}}}
	\subfloat[$\#4$\label{flow_transitions_if_corr_path4l}]{\resizebox{0.25\textwidth}{!}{\input{slug2disp4_if_corr_fig2}}}\\
	\caption {Results for path 4 (Intermittent to Dispersed) : (a,b,c,d) Phases distribution at $t=30~\second$  (not at scale). (e,f,g,h) Evolution of the volume fractions at the outlet with time. (i,j,k,l) Evolution of the average pressure drop at the outlet with time.}
	\label{flow_transitions_if_corr_path4}
\end{figure}

Using the same geometry, fluid properties and flow map as in the previous section, we picked four different paths from one flow pattern to the other. For each path, we simulated four different cases that are marked with different symbols on Figure~\ref{flow_map2} and the corresponding parameters are shown on Table~\ref{tab:case_prop2}. All cases were simulated in 2D with a particle resolution of $L/\Delta r = 312$ which corresponds approximately to $10000$ particles. The simulation time was $30~\second$. At the inlet, each phase is injected with a constant velocity corresponding to its superficial velocity $u_{g,l}^s = \alpha_{g,l} u_{g,l}$. At the outlet, a constant pressure equal to the background pressure is prescribed. The final two-flow patterns for the different cases, the evolution of the volume fraction with time at the outlet and the average pressure drop between both ends of the pipe are shown on Figures~\ref{flow_transitions_if_corr_path1} to \ref{flow_transitions_if_corr_path4}.

\begin{table}[bthp]
	\begin{adjustbox}{max width=\textwidth}
	\begin{tabular}{llllll}
		\toprule
		\textbf{Property} & \textbf{Path 1}   & \textbf{Path 2} & \textbf{Path 3} & \textbf{Path 4} & \textbf{Units}\\
		\midrule
		\textbf{Flow pattern} & \specialcell{Stratified to\\Mist} & \specialcell{Mist to\\Intermittent} & \specialcell{Stratified to\\Intermittent} & \specialcell{Intermittent to\\Dispersed} & -\\
		\textbf{Superficial velocities $\#1$} ($u_g^s$,$u_l^s$) & ($1.0$,$0.06$) & ($3.0$,$0.4$) & ($0.3$,$0.15$) & ($0.22$,$3.0$) & $\meter\per\second$\\
		\color{white}{\textbf{Superficial velocities}} \color{black}{\textbf{$\#2$}} \color{white}{($u_g^s$,$u_l^s$)} & ($1.5$,$0.06$) & ($2.0$,$0.8$) & ($0.25$,$0.25$) & ($0.15$,$4.0$) & $\meter\per\second$\\
		\color{white}{\textbf{Superficial velocities}} \color{black}{\textbf{$\#3$}} \color{white}{($u_g^s$,$u_l^s$)} & ($2.5$,$0.06$) & ($1.0$,$1.2$) & ($0.2$,$0.6$) & ($0.1$,$6.0$) & $\meter\per\second$\\
		\color{white}{\textbf{Superficial velocities}} \color{black}{\textbf{$\#4$}} \color{white}{($u_g^s$,$u_l^s$)} & ($3.5$,$0.06$) & ($0.5$,$1.6$) & ($0.2$,$1.0$) & ($0.07$,$7.0$) & $\meter\per\second$\\
		\textbf{Reynolds number $\#1$} ($Re=\frac{(u_g^s+u_l^s)D}{\nu_g \alpha_g + \nu_l \alpha_l}$) & $302.86$ & $971.43$ & $128.57$ & $920.00$ & -\\
		\color{white}{\textbf{Reynolds number}} \color{black}{\textbf{$\#2$}} \color{white}{($Re=\frac{(u_g^s+u_l^s)D}{\nu_g \alpha_g + \nu_l \alpha_l}$)} & $445.71$ & $800.00$ & $142.86$ & $1185.71$ & -\\
		\color{white}{\textbf{Reynolds number}} \color{black}{\textbf{$\#3$}} \color{white}{($Re=\frac{(u_g^s+u_l^s)D}{\nu_g \alpha_g + \nu_l \alpha_l}$)} & $731.43$ & $228.57$ & $642.86$ & $1742.86$ & -\\
		\color{white}{\textbf{Reynolds number}} \color{black}{\textbf{$\#4$}} \color{white}{($Re=\frac{(u_g^s+u_l^s)D}{\nu_g \alpha_g + \nu_l \alpha_l}$)} & $1017.14$ & $600.00$ & $342.86$ & $2020.00$ & -\\
		\bottomrule
	\end{tabular}
\end{adjustbox}
	\caption{Cases properties for the different paths}
	\label{tab:case_prop2}
\end{table}

On Figures~\ref{flow_transitions_if_corr_path1a} to~\ref{flow_transitions_if_corr_path1d}, the transition from a stratified to a mist flow (path 1) is shown. As the superficial velocity of the light phase increases from case $\#1$ to case $\#4$, the volume fraction of the light phase is going to $\approx 1$ while the volume fraction of the heavy fluid is going to $0$. In between, we can see that the heavy fluid layer is divided in pieces until being completely dispersed by the light phase. Volume fractions plots of Figures~\ref{flow_transitions_if_corr_path1e} to~\ref{flow_transitions_if_corr_path1h} show that the transition between the two patterns goes through a phase of intermittent flow where the heavy phase layer dislocates forming drops. On Figures~\ref{flow_transitions_if_corr_path1i} to~\ref{flow_transitions_if_corr_path1l}, the pressure drop level increases from $\approx 10~\pascal$ to $\approx 150~\pascal$ and present stronger and stronger oscillations while transitioning to the mist flow pattern.

The transition from mist flow to intermittent flow (path 2) is presented on Figures~\ref{flow_transitions_if_corr_path2a} to~\ref{flow_transitions_if_corr_path2d}. From a quasi-mist flow in case $\#1$, we see that, as the heavy phase superficial velocity increases, the fluid becomes wavy and then evolves towards an intermittent flow. This is magnified by the volume fractions evolution plots of Figures~\ref{flow_transitions_if_corr_path2e} to~\ref{flow_transitions_if_corr_path2h} where we see that the light phase dominant in case $\#1$ whereas the heavy phase is dominant in case $\#4$. In all four cases, volume fractions are showing significant oscillations so that it is not obvious to qualify where the flow becomes really intermittent. This supports the well-known fact that flow maps are only an indicative tool and that the transition lines are not lines but smooth transitions areas. Concerning the pressure drop evolution of Figures~\ref{flow_transitions_if_corr_path2i} to~\ref{flow_transitions_if_corr_path2l}, we can see that the average level drops from $\approx 125 ~\pascal$ to $\approx 25 ~\pascal$. The case $\#3$ is particular since in that case, superficial velocities and volume fractions are the same, therefore it is a very stable case where the pressure drop oscillations are the smallest.

On Figures~\ref{flow_transitions_if_corr_path3a} to~\ref{flow_transitions_if_corr_path3d}, we present the phases distribution for the transition from a stratified flow pattern to an intermittent flow pattern. As the heavy phase velocity increases, the heavy phase becomes dominant and the interface with the light phase is more and more wavy near the outlet which prefigures the emergence of an intermittent flow. Volume fractions plots shown on Figures~\ref{flow_transitions_if_corr_path3e} to~\ref{flow_transitions_if_corr_path3h} support that observation as the amplitudes of their oscillations are increasing when we move towards the intermittent flow pattern. On Figures~\ref{flow_transitions_if_corr_path3i} to~\ref{flow_transitions_if_corr_path3l}, one can note that the pressure drops evolutions behave similarly and their average level increases from $\approx 4 ~\pascal$ to $\approx 8 ~\pascal$.

The transition from intermittent flow to dispersed flow is presented on Figures~\ref{flow_transitions_if_corr_path4a} to~\ref{flow_transitions_if_corr_path4d}. The flow pattern evolves from a disturbed intermittent flow that we could qualify of plug flow towards a dispersed flow as the heavy phase velocity increases. On Figures~\ref{flow_transitions_if_corr_path4e} to~\ref{flow_transitions_if_corr_path4h}, on can observe that volume fractions time series are initially very unstable which is characteristic of an intermittent flow. The light phase and heavy phase volume fractions are going to $ \approx 0$ and $\approx 1$ respectively as the dispersed flow pattern emerges. Concerning the pressure drops of Figures~\ref{flow_transitions_if_corr_path4i} to~\ref{flow_transitions_if_corr_path4l}, as expected, it increases from $\approx 75 ~\pascal$ to $\approx 400 ~\pascal$ with growing oscillations.

To conclude, we have explored the flow map of Figure~\ref{flow_map2} by simulating several cases located around the transitions from one pattern to the other. We observed that the transition areas are not lines but in fact smooth bands. Also, it appears that the intermittent flow area gathers different patterns such as wavy flows (in the lower part), slug flows (in the center) and even plug flows (in the upper area). The pressure drop plots are providing useful information on the pressure field but are showing strong variations which are due to the use of the weakly compressible formulation. This approach is known to generate disturbances in the pressure calculations because of density and pressure are linked through an equation of state~\eqref{tait_eos}. A truly incompressible SPH formulation would likely improve this aspect. See~\cite{hu2007,kunz2015} for multiphase incompressible SPH models. Besides, the inlet, outlet and wall boundary conditions that we have used are known to introduce spurious waves in the flow. We believe that implementing more accurate boundary conditions based on analytical considerations~\cite{Ferrand2012} and adapted to inlet/outlet for multiphase flows could also improve the quality of the results. Nevertheless, we think that these results show that SPH could be a complementary tool to study the emergence of intermittent flow patterns in pipes in industrial applications.

\subsection{Applied cases with high density and viscosity ratios}\label{oilgas}

\begin{figure}[bthp]
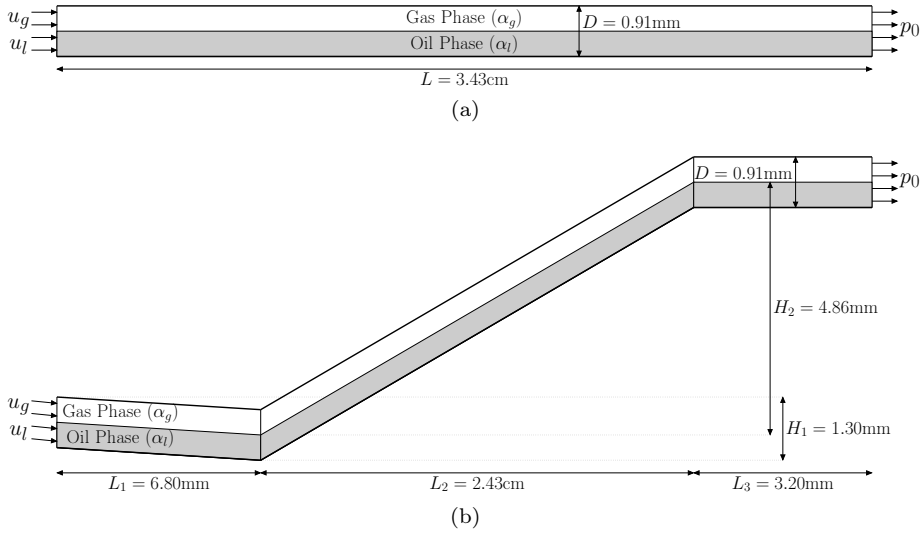

	\begin{center}
		\subfloat[]{\resizebox{1.0\textwidth}{!}{\input{applied_cases_geometry1_micro}}}\\
		\subfloat[]{\resizebox{1.0\textwidth}{!}{\input{applied_cases_geometry2_micro}}}
		\caption {Geometrical configurations for the two applied cases (not at scale). (a) Hydrodynamic slugging. (b) Terrain slugging.}
		\label{init_geo2}
	\end{center}
\end{figure}

\begin{figure}[bthp]
	\begin{center}
		\includegraphics[width=1.0\textwidth,height=0.04\textwidth]{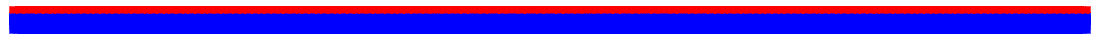}\\
		\includegraphics[width=1.0\textwidth,height=0.04\textwidth]{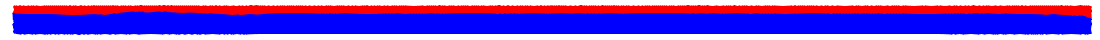}\
		\includegraphics[width=1.0\textwidth,height=0.04\textwidth]{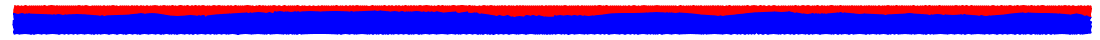}\\
		\includegraphics[width=1.0\textwidth,height=0.04\textwidth]{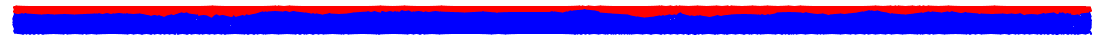}\\
		\includegraphics[width=1.0\textwidth,height=0.04\textwidth]{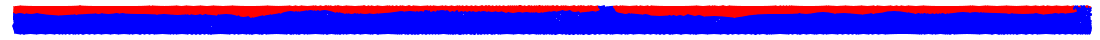}\\
		\includegraphics[width=1.0\textwidth,height=0.04\textwidth]{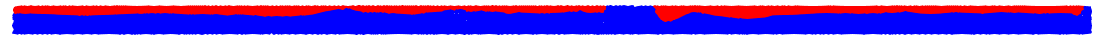}\\
		\includegraphics[width=1.0\textwidth,height=0.04\textwidth]{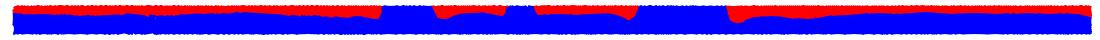}\\
		\includegraphics[width=1.0\textwidth,height=0.04\textwidth]{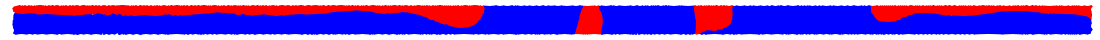}\\
		\includegraphics[width=1.0\textwidth,height=0.04\textwidth]{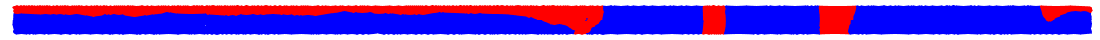}\\
		\includegraphics[width=1.0\textwidth,height=0.04\textwidth]{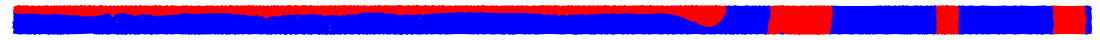}\\
		\caption {Phases distribution for the hydrodynamic slugging case (not at scale). From top to bottom, at $t=0~\second$, $0.01~\second$, $0.02~\second$, $0.025~\second$, $0.026~\second$, $0.0275~\second$, $0.03~\second$, $0.035~\second$, $0.04~\second$ and $0.05~\second$}
		\label{hs}
	\end{center}
\end{figure}

\begin{figure}[bthp]
	\begin{center}
		\subfloat[]{ \makebox[\textwidth][c]{\includegraphics[width=1.2\textwidth]{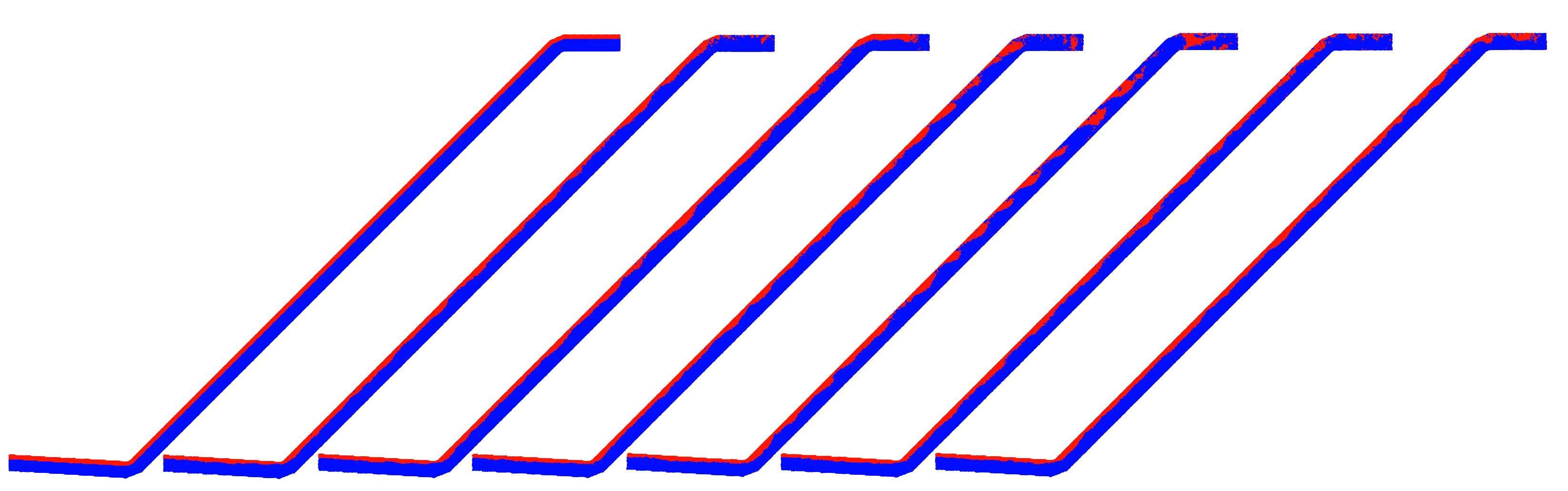}}}\\
		\subfloat[\label{volfracts}]{\resizebox{0.5\textwidth}{!}{\input{terrain_slugging_fig1_2}}}
		\subfloat[\label{fftts}]{\resizebox{0.5\textwidth}{!}{\input{terrain_slugging_fig7_3}}}
		\caption {(a) Phases distribution for the terrain slugging case. From left to right, at $t=0~\second$, $0.05~\second$, $0.1~\second$, $0.13~\second$, $0.188~\second$, $0.2~\second$ and $0.25~\second$ (b) Evolution of the volume fractions at the outlet with time. (c) Fourier transform of the gas phase volume fraction time series (in red on (b))}
		\label{ts}
	\end{center}
\end{figure}

In order to further demonstrate the potential of SPH to model intermittent flows, we have simulated two applied cases. The fluids considered for these cases are generic oil and natural gas whose properties are indicated in Table~\ref{tab:mat_prop3}. The two different geometries corresponding to a hydrodynamic slugging case and a terrain slugging case are shown on Figure~\ref{init_geo2} and are discretized with resolutions $L/\Delta r = 634$ ($\approx 10000$ particles) and $L/\Delta r = 687$ ($\approx 15000$ particles) respectively. Simulation times are $0.05\second$ and $0.25\second$ respectively. Fluids are injected with superficial velocities and volume fractions given in Table~\ref{tab:case_prop3}. Note that we have chosen to work with a micro-geometry for computational and time constraints while trying to preserve realistic proportions. Phases distributions for both cases are shown on Figures~\ref{hs} and~\ref{ts}.

\begin{table}[bthp]
	\centering
	\begin{adjustbox}{max width=\textwidth}
		\begin{tabular}{llll}
			\toprule
			\textbf{Property}         & \textbf{Gas Phase}   & \textbf{Oil Phase} & \textbf{Units}\\
			\midrule
			\textbf{Density} ($\rho$) & $1.6454$         & $702.6926$          & $\kilogram\per\meter^{3}$\\
			\textbf{Viscosity} ($\mu$)& $1.27\times 10^{-5}$ & $4.19\times 10^{-4}$  & $\pascal .\second$\\
			\textbf{Adiabatic index} ($\gamma$) & $1.4$ & $7$  & -\\
			\textbf{Surface Tension} ($\sigma^{nw}$) & \multicolumn{2}{l}{0.02139} & $\newton\per\meter$\\
			\textbf{Contact Angle} ($\theta_c$) & \multicolumn{2}{l}{90} & $\degree$\\
			\textbf{Gravity} ($g_z$) & \multicolumn{2}{l}{-9.81} & $\meter\per\second^2$\\
			\bottomrule
		\end{tabular}
	\end{adjustbox}
	\caption{Physical Properties}
	\label{tab:mat_prop3}
\end{table}

\begin{table}[bthp]
	\begin{adjustbox}{max width=\textwidth}
		\begin{tabular}{llllll}
			\toprule
			\textbf{Property} & \textbf{Hydrodynamic slugging}   & \textbf{Terrain slugging} & \textbf{Units}\\
			\midrule
			\textbf{Volume fractions} ($\alpha_g$,$\alpha_l$) & ($0.25$,$0.75$) & ($0.25$,$0.75$) & $\meter\per\second$\\
			\textbf{Superficial velocities} ($u_g^s$,$u_l^s$) & ($0.35$,$0.76125$) & ($0.35$,$0.76125$) & $\meter\per\second$\\
			\bottomrule
		\end{tabular}
	\end{adjustbox}
	\caption{Cases Properties}
	\label{tab:case_prop3}
\end{table}

On Figure~\ref{hs}, it is possible to observe the typical formation process of a hydrodynamic slug. First, from $t=0~\second$ to $t=0.025~\second$, waves begin to grow. At $t=0.026~\second$, one the waves' crest is high enough to reach the top of pipe : a slug is formed. From $t=0.0275~\second$ to $t=0.03~\second$, other waves reach the top of the pipe forming new slugs. After that, one can note that some slugs see their length reduced and their height increased until occupying the whole pipe's height. This example confirms that SPH can reproduce the dynamics of slug flows with high viscosity and density ratios (here $\approx 32$ and $\approx 427$ respectively).

On Figure~\ref{ts}, we present the results of a "riser-like" case where the oil and gas mixture extracted from the reservoir is lifted from the sea ground to the land. A slug flow does not have the required distance to grow in the initial descending part of the pipe. However, under the effect of gravity, slugging begins to occur in the ascending part of the pipe. The evolution of volume fractions at the outlet of Figure~\ref{volfracts} shows strong oscillations as expected in a slug flow. When performing a Fourier transform analysis on the gas phase volume fraction evolution as presented on Figure~\ref{fftts}, one can see that one frequency clearly dominates, thus suggesting that the slug frequency in that particular case geometry would be around $226~\hertz$.

\section{Conclusion}
In this paper, we have combined ideas from~\cite{tafuni2018,alvaro2017} to present inlet/outlet boundary conditions for multiphase flows in SPH. These boundary conditions have then been used to simulate intermittent flows in pipes. First, we have shown that our SPH implementation is able to reproduce  four different flow regimes (mist flow, dispersed flow, intermittent flow and stratified flow) predicted by Taitel and Dukler's flow map. Then, we focused on the transition processes from one flow pattern to another : stratified to mist, stratified to intermittent, mist to intermittent and intermittent to dispersed. 

From theses simulations, we have confirmed that SPH has a strong potential to model two-phase flow in pipes and to understand how one pattern evolves into another one. However, we believe that our boundary conditions, although simple and easy to implement, are not optimal and that a semi-analytic approach~\cite{Ferrand2012} could improve the quality of the results. Similarly, pressure drops of our simulations have reasonable levels that correspond to flow velocities but are also presenting large oscillations due to the weakly compressible approach. A truly incompressible formulation based on a Poisson solver could remediate this issue. Finally, we have simulated two cases of hydrodynamic slugging and terrain slugging involving high density and viscosity ratios to demonstrate the applicability of multiphase SPH to more realistic problems.

Albeit satisfactory, our results could certainly be improved. To this end, three main tasks can be identified. First, it would be interesting to verify if including the turbulence effects and increasing the number of particles would stabilize the pressure field, especially for high Reynolds cases. Then, it is needed to clarify the influence of shifting and interface correction on the results through a parameter study. Finally, a comparison with results obtained from another numerical method or from a commercial software would be mandatory to further assess the potential interest of SPH in slugging modeling. In addition, recently proposed multiphase SPH models such as~\cite{hammani2018} could possibly enhance the quality of the simulations.

Although in general slower than other numerical methods, this study intended to show that SPH, thanks to its ease of handling interface dynamics, could be used in industrial applications to model intermittent flows and in particular slug and plug flows with reliable results. This work hopes to serve as a basis on which to build more complex SPH models including turbulence effects to capture the flow behavior with more accuracy.

\section*{Acknowledgments}
	The authors would like to thank Total for its scientific and financial support. The authors would also like to thank Alexandre Boucher for many fruitful discussions on the topic.

\section*{References}
\bibliographystyle{abbrv}
\bibliography{bibliography/references}

\appendix

\section{Validation Cases}\label{appendix}

For all the following validation cases, no kernel gradient correction or shifting or interface correction were used.

\subsection{Lid-driven Cavity Flow}
\begin{figure}[bthp]
	\begin{center}
		\resizebox{0.6\textwidth}{!}{\input{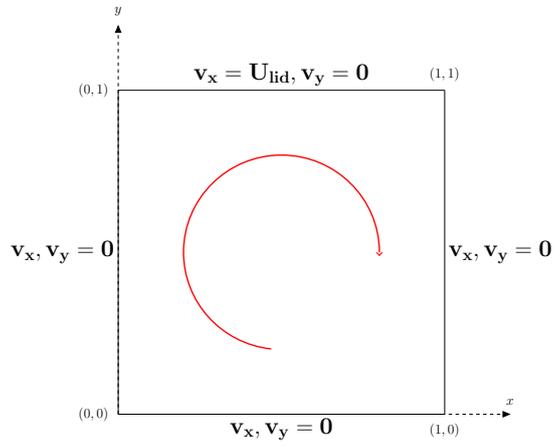}}
		\caption {The 2D Lid-driven Cavity Flow problem}
		\label{cavity_flow}
	\end{center}
\end{figure}

The goal of this section is to validate the implementation of SPH for the single phase Navier-Stokes case. The test case chosen for this purpose is the well-known 2D lid-driven cavity flow problem shown on Figure~\ref{cavity_flow}. This is a common problem in the fluid mechanics community and numerous reference solutions performed with different numerical methods are available in the literature. In this case, we use Ghia et al. solution as a reference~\cite{ghia1982}.

The Reynolds number for this problem is defined as follows $Re = \frac{U_{\text{lid}} L}{\nu}$ where $U_{\text{lid}}$ is the velocity of the imposed at the top boundary, $\nu$ is the kinematic viscosity and $L$ is the characteristic length of the problem. The simulations were performed for $Re=100$ and $1000$ and for $50\times 50$, $100\times 100$ and $200\times 200$ particles. The density is set to $1000~\kilogram\per\meter^{3}$, the velocity of the lid is $U_{\text{lid}}=1~\meter\per\second$, the domain is $L_x \times L_y = 1~\meter\times 1~\meter$ and the viscosity $\nu$ is adjusted to reach the desired Reynolds number. $\gamma=7.0$ for both fluids.

The velocity boundary condition at the top boundary has been applied using the procedure described in section \ref{sec:bc}. For the other boundaries, a no-slip boundary conditions has been applied. The simulations are terminated when a steady state is reached (i.e. $\sqrt{\sum_a\frac{\lvert\rho_a^{n+1}-\rho_a^{n}\rvert}{\rho_a^{n}}}< 1e^{-2}$ or after $60~\second$ of real simulated time).

\subsubsection{$Re=100$}

\begin{figure}[bthp]
	\captionsetup[subfigure]{labelformat=empty}
	\centering
	\subfloat[] {\includegraphics[width=0.06\textwidth]{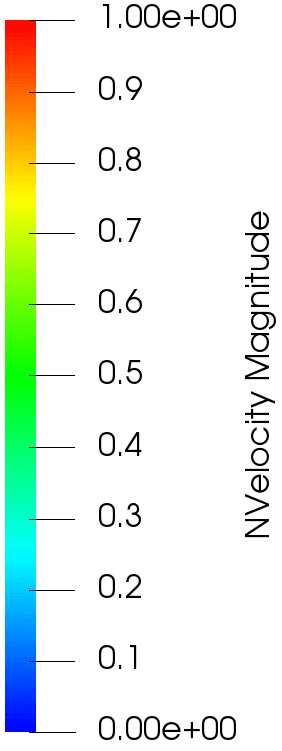}}
	\subfloat[$50\times 50$] {\includegraphics[width=0.225\textwidth]{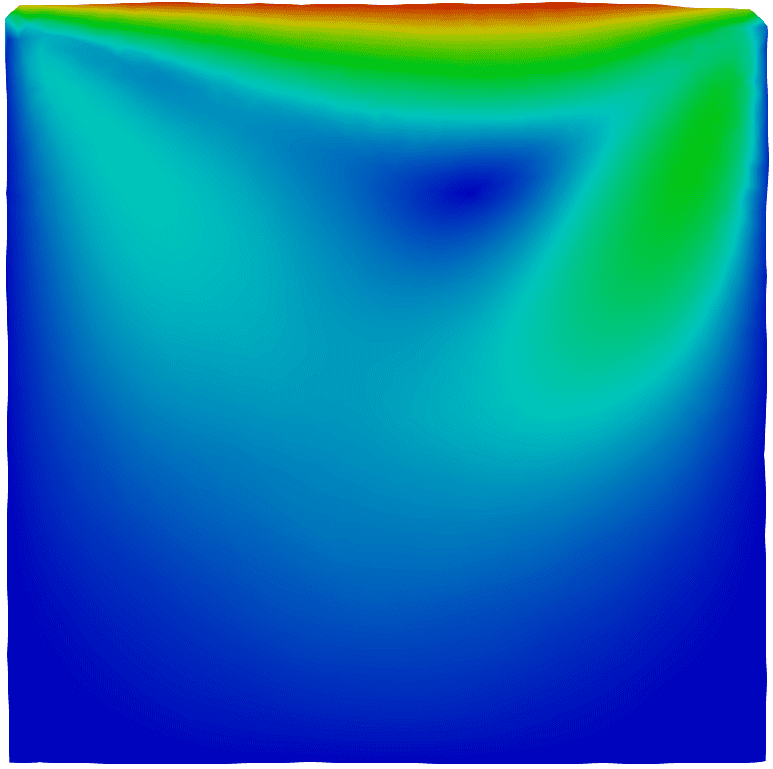}}\hspace{0.5cm}
	\subfloat[] {\includegraphics[width=0.06\textwidth]{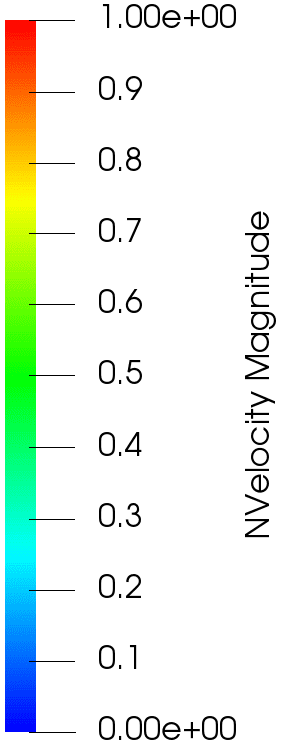}}
	\subfloat[$100\times 100$] {\includegraphics[width=0.23625\textwidth]{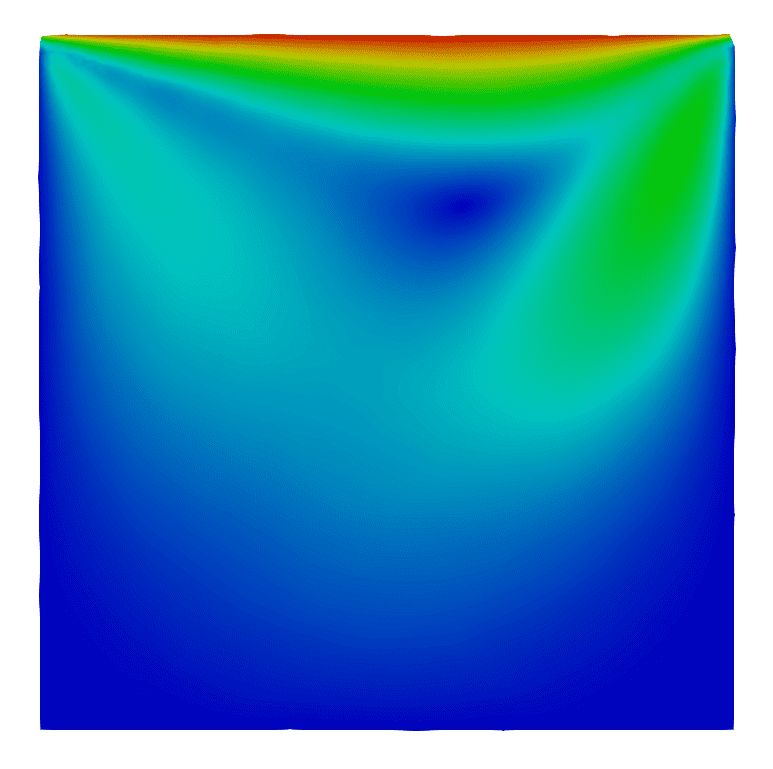}}\hspace{0.5cm}
	\subfloat[] {\includegraphics[width=0.06\textwidth]{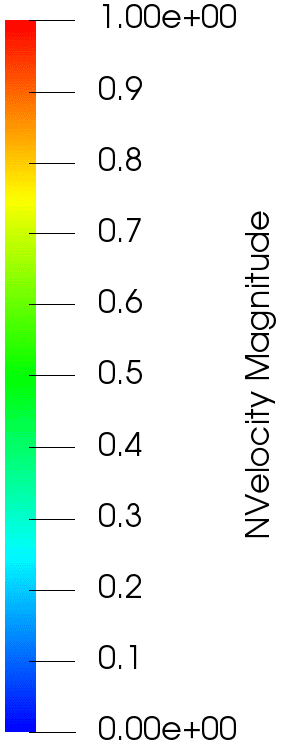}}
	\subfloat[$200\times 200$] {\includegraphics[width=0.225\textwidth]{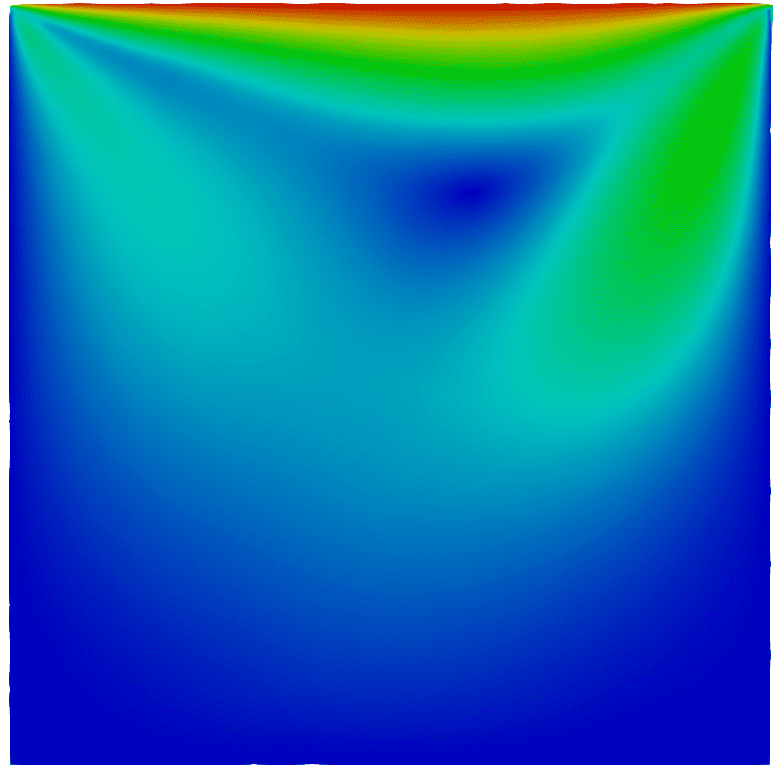}}\\
	\hspace{0.45cm}
	\subfloat[$50\times 50$ ] {\includegraphics[width=0.225\textwidth]{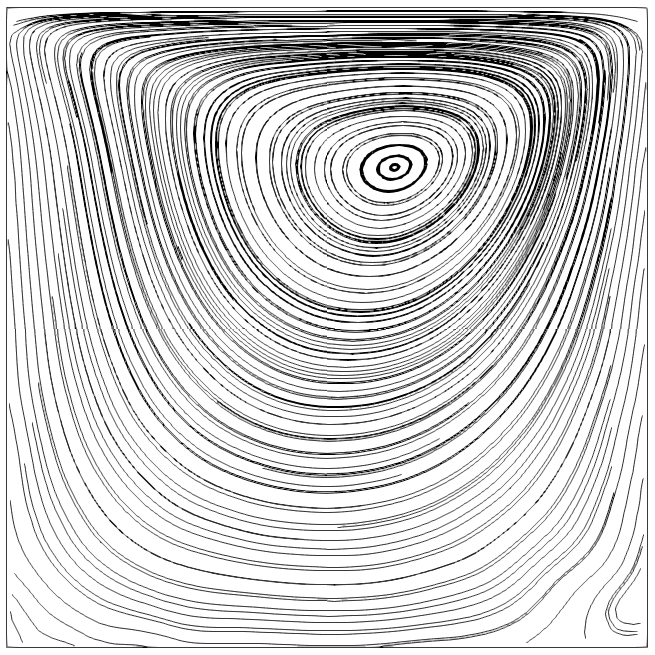}}\hspace{1.0cm}
	\subfloat[$100\times 100$ ] {\includegraphics[width=0.23625\textwidth]{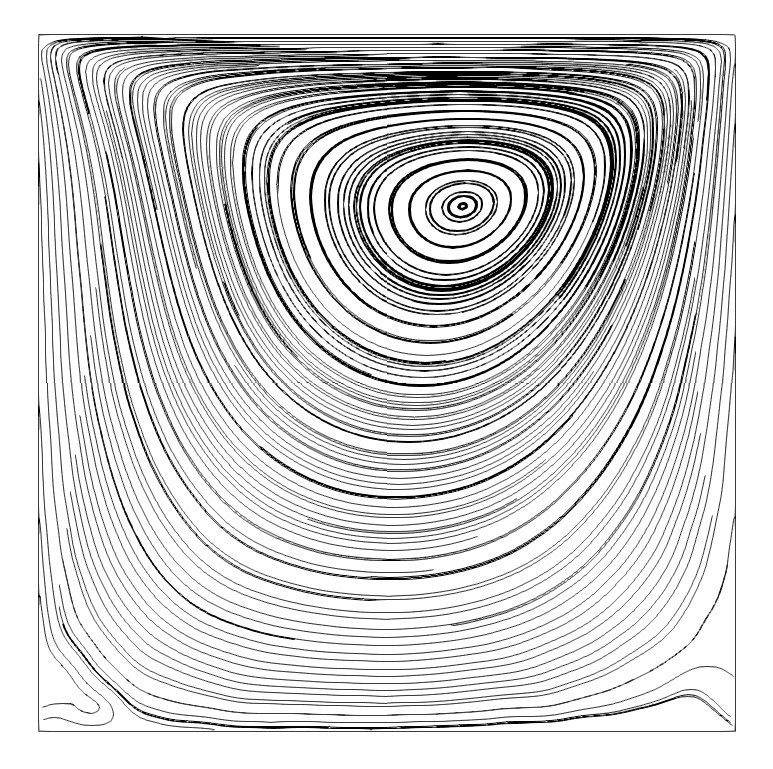}}\hspace{1.0cm}
	\subfloat[$200\times 200$] {\includegraphics[width=0.225\textwidth]{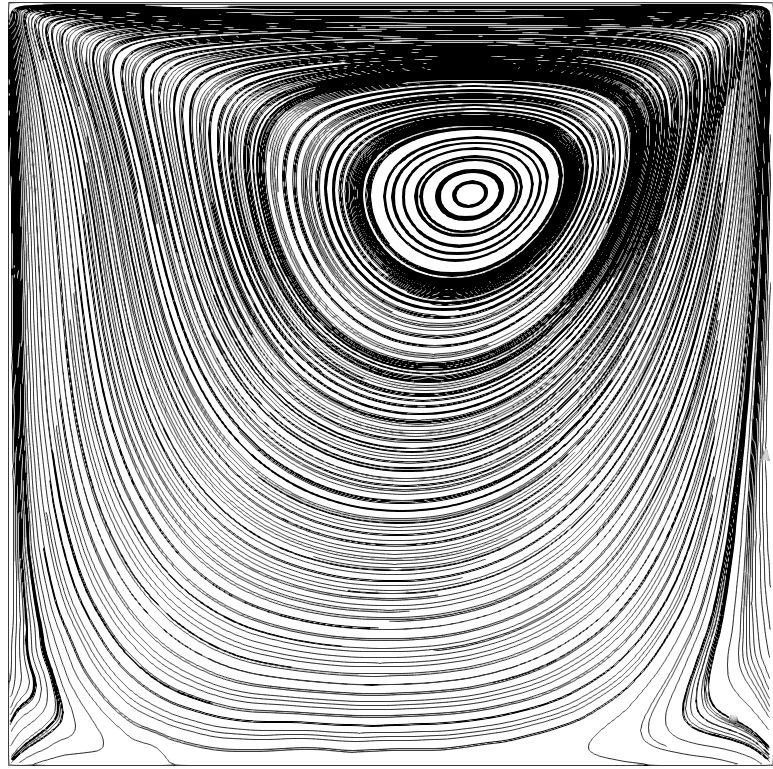}}
	\caption {Steady-state results for $Re=100$ showing the velocity distribution (left) and the corresponding streamlines (right)}
	\label{re100_sph}
\end{figure}

\begin{figure}[bthp]
	\begin{center}
		\captionsetup[subfigure]{labelformat=empty}
		\subfloat[]{\resizebox{0.4\textwidth}{!}{\input{Re100}}}
		\subfloat[]{\resizebox{0.4\textwidth}{!}{\input{LDCF_SPH_loglog_Re100_Y}}}\\
		\subfloat[]{\resizebox{0.4\textwidth}{!}{\input{Re100_2}}}
		\subfloat[]{\resizebox{0.4\textwidth}{!}{\input{LDCF_SPH_loglog_Re100_X}}}
		\caption {$Re=100$}
		\label{re_100}
	\end{center}
\end{figure}

When $Re=100$, SPH is able to reproduce the velocity field accurately. The maximum error is reached at the boundaries but remains low ($\lesssim 2 \%$ for the $200\times 200$ case close to the boundaries, figure not shown here). On Figure~\ref{re_100}, the $L_2$ norm of the error with respect to the reference solution has been plotted in function of the number of particles in log scale. We observe a linear "mesh" convergence for the both velocity fields when the particle resolution increases. The convergence rate is $\approx 0.1$.

Concerning the spatial distribution of the flow, the theory predicts the appearance of two vortexes at the two bottom corners of the domain. As shown on Figure~\ref{re100_sph}, SPH is not able to reproduce those two vertexes but instead has flow perturbations in the concerned areas.

    \begin{figure}[bthp]
	\captionsetup[subfigure]{labelformat=empty}
	\centering
	\subfloat[$50\times 50$ - $t=2.48s$] {\includegraphics[width=0.225\textwidth]{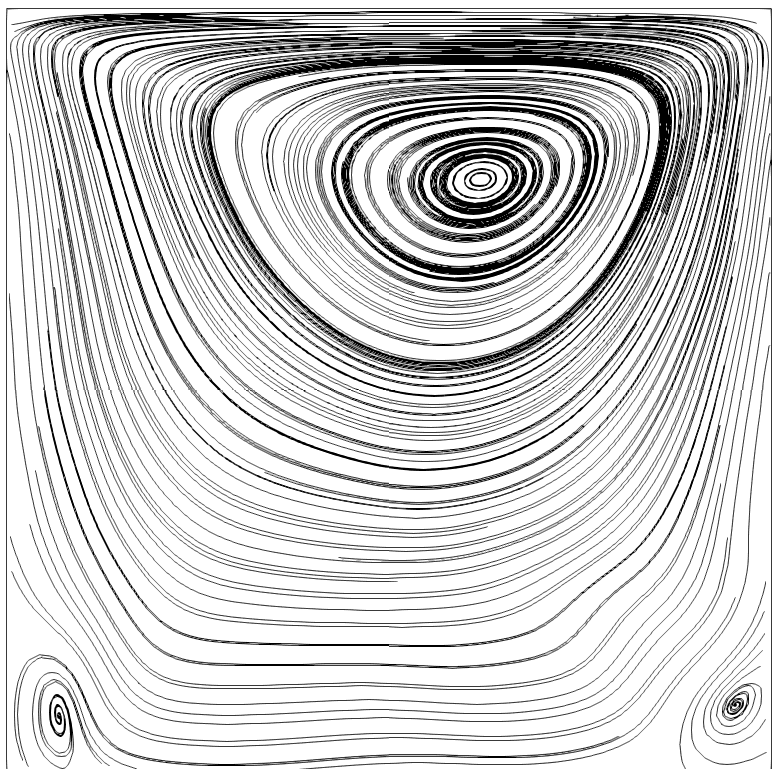}}\hspace{0.5cm}
	\subfloat[$100\times 100$ - $t=46.66s$] {\includegraphics[width=0.225\textwidth]{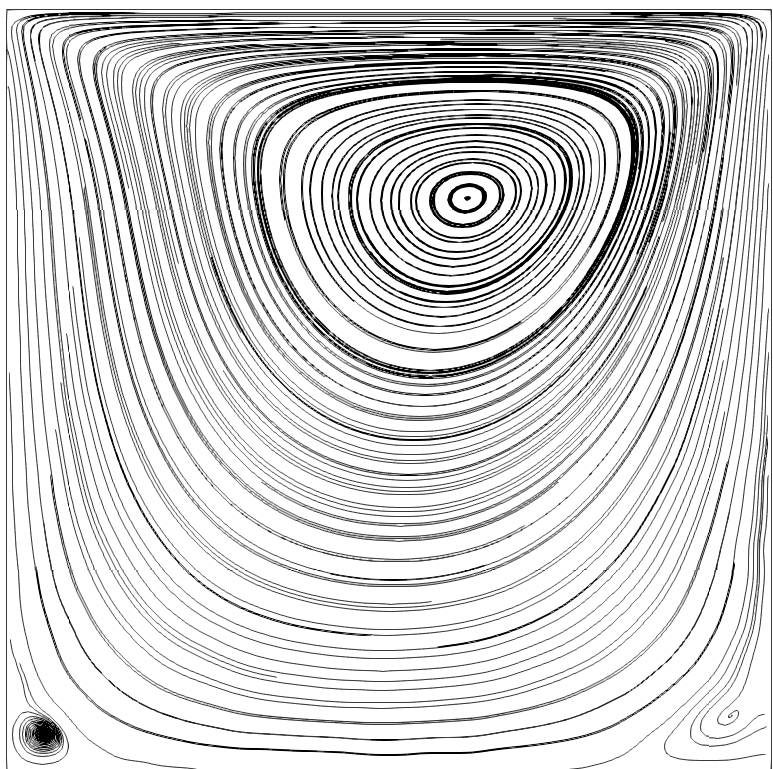}}\hspace{0.5cm}
	\subfloat[$200\times 200$ - $t=54.31s$] {\includegraphics[width=0.225\textwidth]{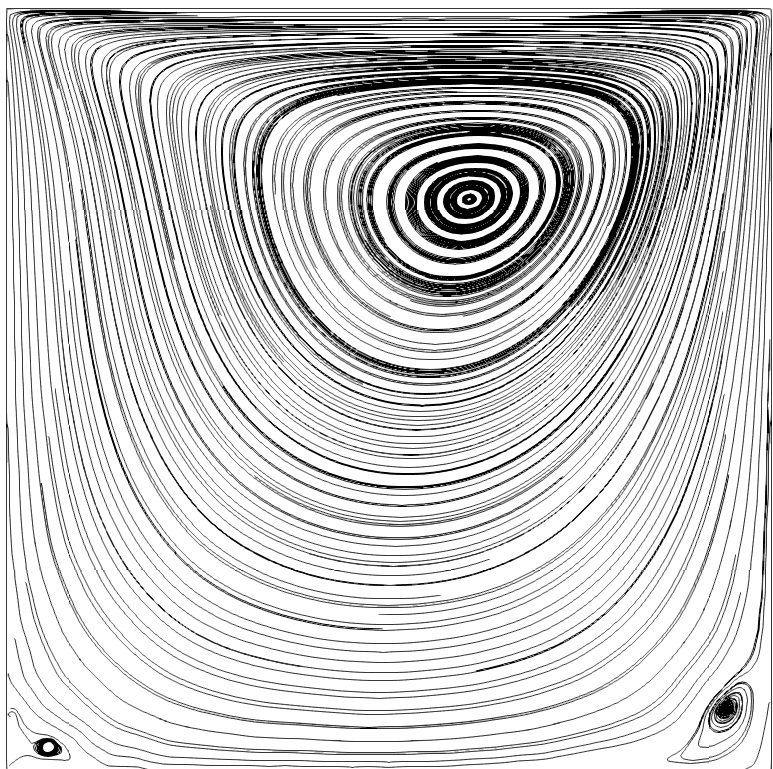}}
	\caption {Streamlines for $Re=100$ at selected timesteps}
	\label{re100_sl_sph}
\end{figure}

On Figure~\ref{re100_sl_sph}, one can note that the two expected vertexes at the corners are in fact appearing during the SPH simulations but they are highly unstable. They keep forming (together or independently) and vanishing as the simulation progresses. It indicates that SPH captures an instability in the correct areas but fails to reach a steady state thus the formation of spurious perturbations. Those vertexes being of small intensity, their formation is likely to be affected by boundary conditions.

\subsubsection{$Re=1000$}

    \begin{figure}[bthp]
	\captionsetup[subfigure]{labelformat=empty}
	\centering
	\subfloat[] {\includegraphics[width=0.06\textwidth]{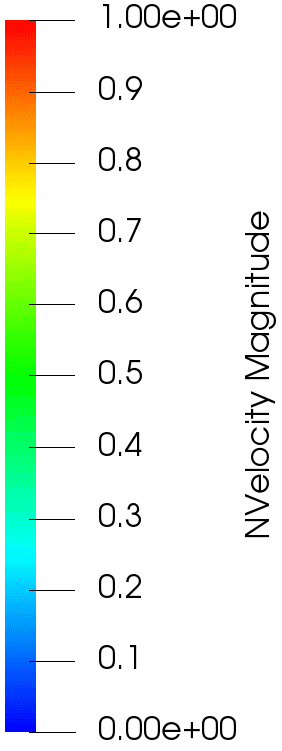}}
	\subfloat[$50\times 50$] {\includegraphics[width=0.225\textwidth]{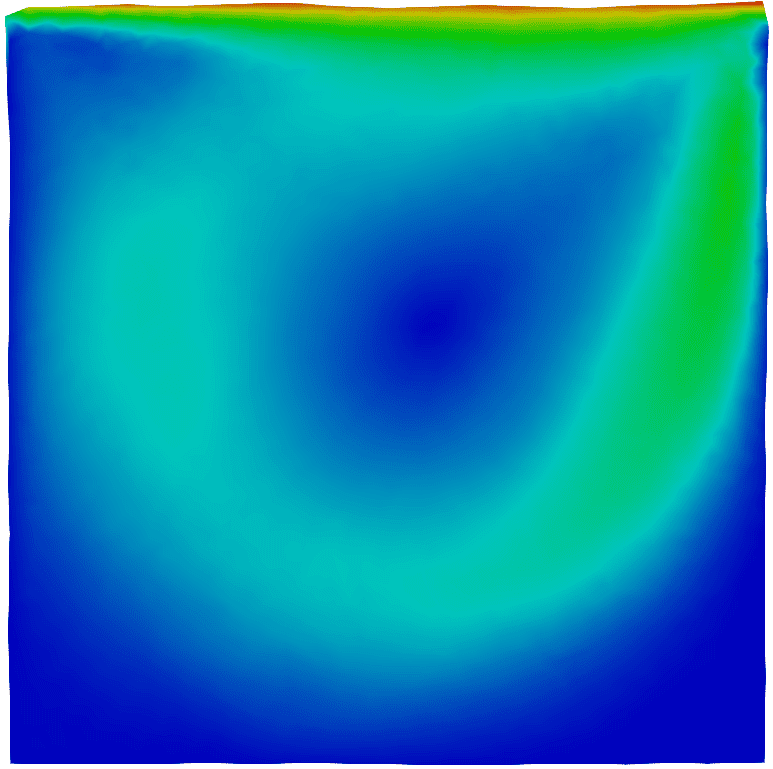}}\hspace{0.5cm}
	\subfloat[] {\includegraphics[width=0.06\textwidth]{SPH_Re1000_40000_VelMag_legend.png}}
	\subfloat[$100\times 100$] {\includegraphics[width=0.225\textwidth]{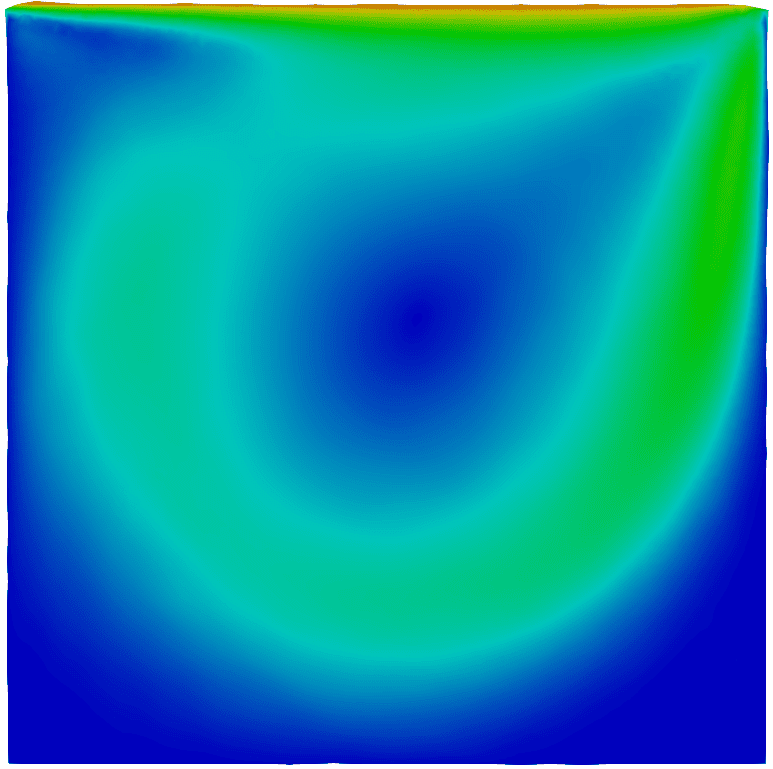}}\hspace{0.5cm}
	\subfloat[] {\includegraphics[width=0.06\textwidth]{SPH_Re1000_40000_VelMag_legend.png}}
	\subfloat[$200\times 200$] {\includegraphics[width=0.225\textwidth]{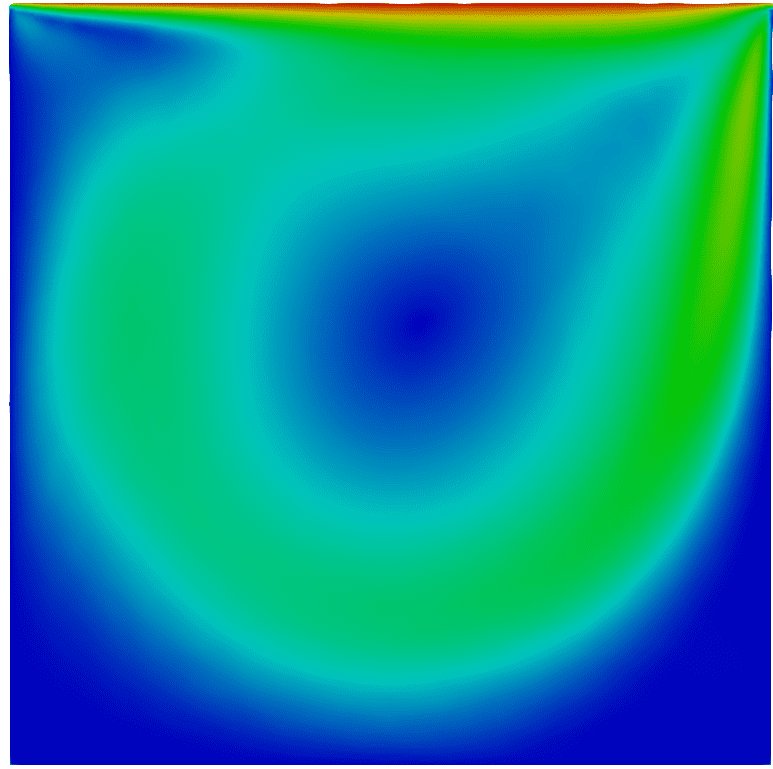}}\\
	\hspace{0.45cm}
	\subfloat[$50\times 50$] {\includegraphics[width=0.225\textwidth]{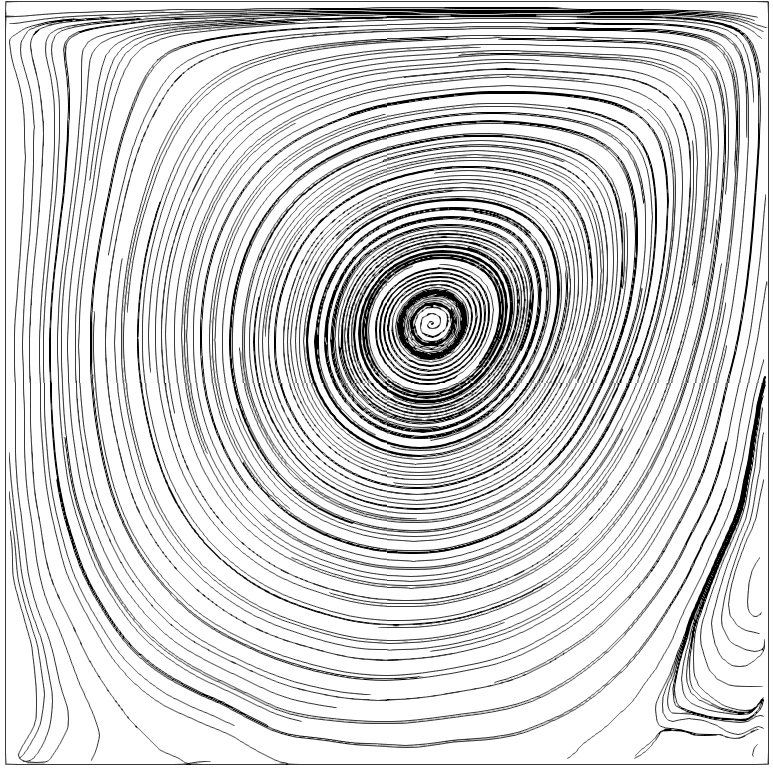}}\hspace{1.0cm} 
	\subfloat[$100\times 100$] {\includegraphics[width=0.225\textwidth]{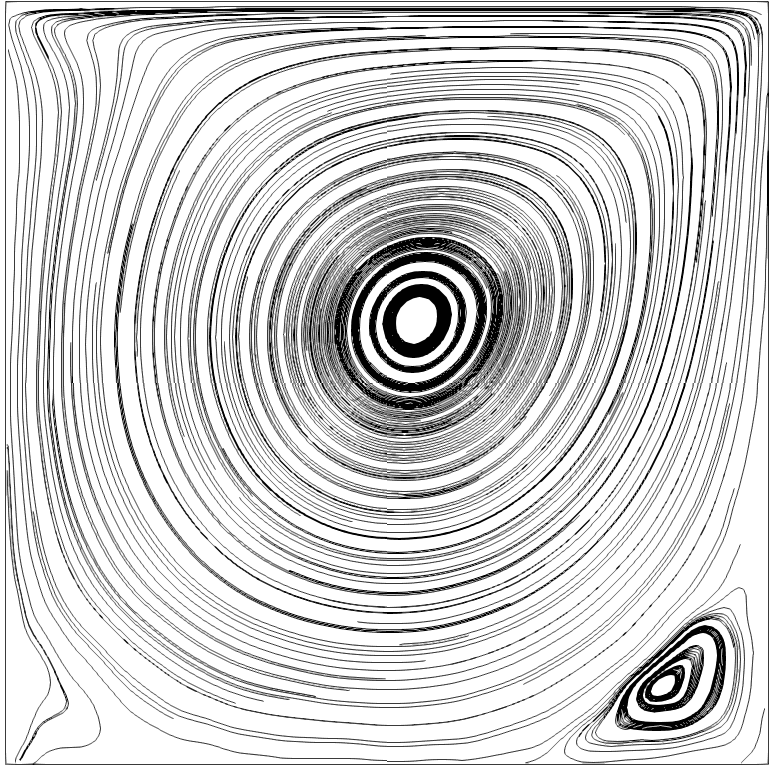}}\hspace{1.0cm}
	\subfloat[$200\times 200$] {\includegraphics[width=0.225\textwidth]{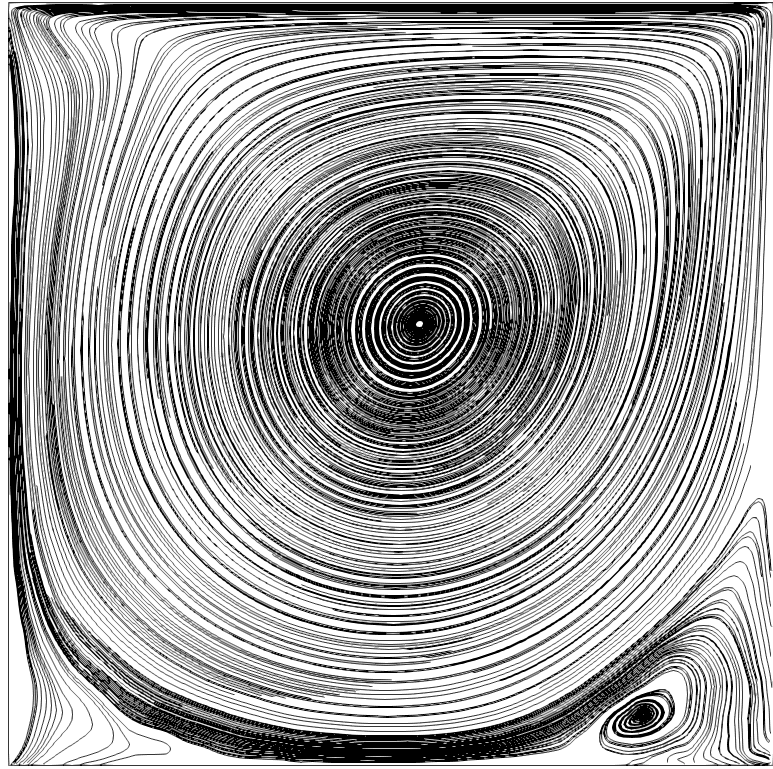}}
	\caption {SPH steady-state results for $Re=1000$ showing the velocity distribution (left) and the corresponding streamlines (right)}
	\label{re1000_sph}
\end{figure}

\begin{figure}[bthp]
	\begin{center}
		\captionsetup[subfigure]{labelformat=empty}
		\subfloat[]{\resizebox{0.4\textwidth}{!}{\input{Re1000}}}
		\subfloat[]{\resizebox{0.4\textwidth}{!}{\input{LDCF_SPH_loglog_Re1000_Y}}}\\		
		\subfloat[]{\resizebox{0.4\textwidth}{!}{\input{Re1000_2}}}
		\subfloat[]{\resizebox{0.4\textwidth}{!}{\input{LDCF_SPH_loglog_Re1000_X}}}		
		\caption {$Re=1000$}
		\label{re_1000}
	\end{center}
\end{figure}

For $Re=1000$, at the highest resolution, SPH gives an error of $\approx 15 \%$ at the right boundary and $\leq 6 \%$ in the interior of the cavity (figure not shown here). As in the previous case, we observe an approximate linear "mesh" convergence for both velocity fields as shown on Figure~\ref{re_1000}. The convergence rate is $\approx 0.25$.

\begin{figure}[bthp]
	\captionsetup[subfigure]{labelformat=empty}
	\centering
	\subfloat[$50\times 50$ - $t=52.83~\second$] {\includegraphics[width=0.225\textwidth]{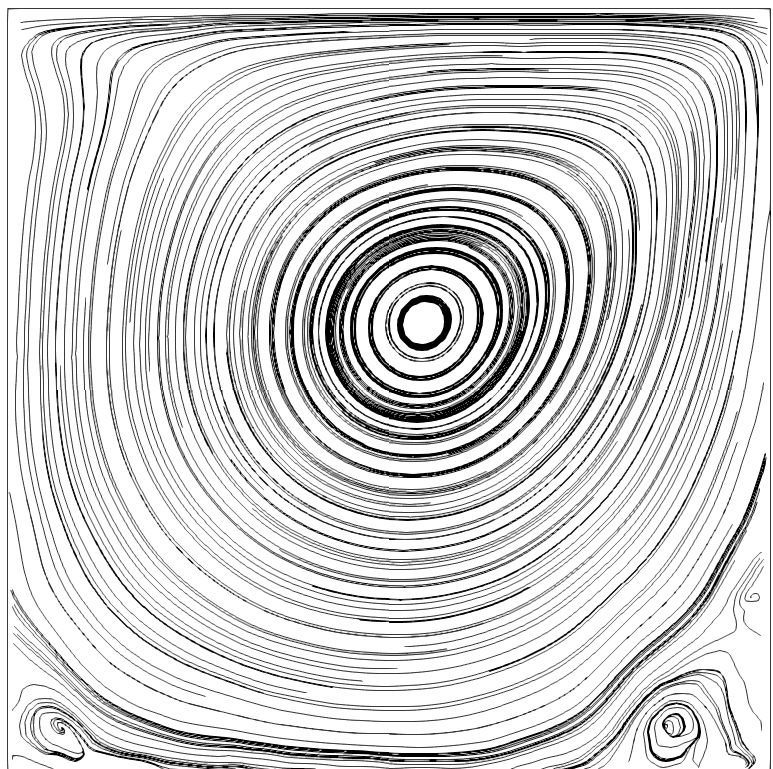}}\hspace{0.5cm}
	\subfloat[$100\times 100$ - $t=50.5~\second$] {\includegraphics[width=0.225\textwidth]{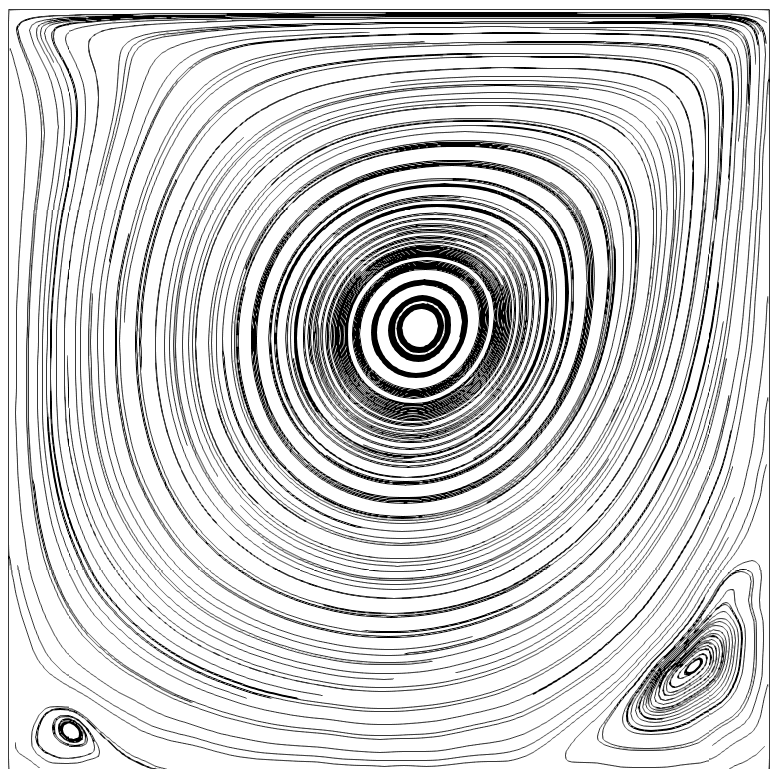}}\hspace{0.5cm}
	\subfloat[$200\times 200$ - $t=52.6~\second$] {\includegraphics[width=0.225\textwidth]{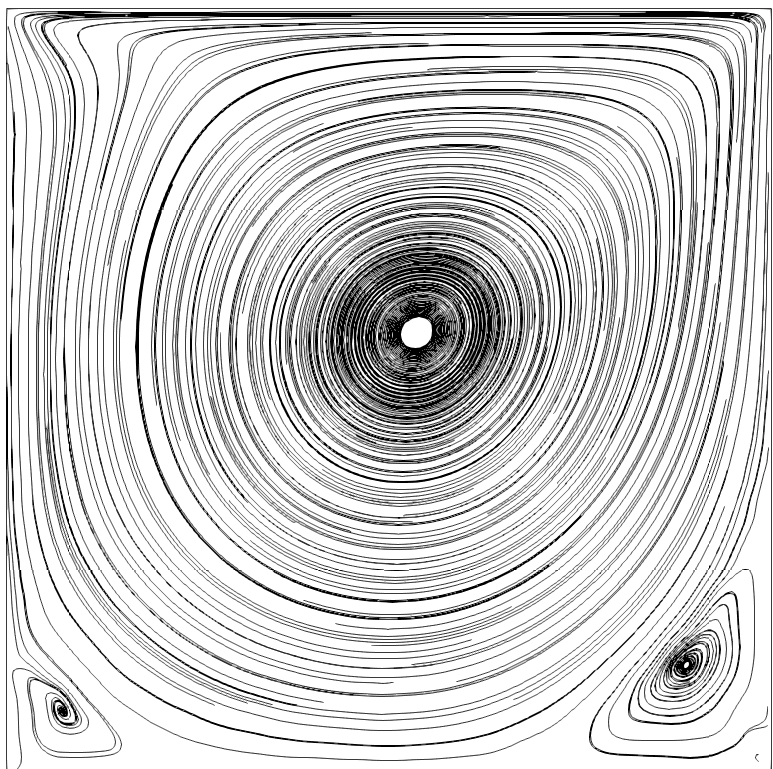}}
	\caption {Streamlines for $Re=1000$ at selected timesteps}
	\label{re1000_sl_sph}
\end{figure}

For this Reynolds number, it can be seen on Figure~\ref{re1000_sph} that SPH is capable of generating a vertex pattern at the bottom right corner for the two highest resolutions but it is unstable for the smallest resolution. Moderate deviations of the flow indicating a potential growing vortex can be seen at the bottom left corner. When computing the streamlines for selected timesteps of the SPH simulations as shown on Figure~\ref{re1000_sl_sph}, it is seen that all three resolutions are generating vertexes in the correct spots but only the $200 \times 200$ case manage to stabilize one at the bottom right corner.

\subsection{Laplace's law}

In this section, the objective is to validate the implementation of the surface tension model described in section~\ref{sec:st}. The standard square to droplet test case is simulated and when a steady state is reached, the pressure difference between the exterior and the interior of the bubble is measured and compared to Laplace's formula

\begin{equation}
\Delta P = \frac{\sigma}{R} = \frac{\sigma \sqrt{\pi}}{a}
\end{equation}

\noindent with $\Delta P$ the pressure difference, $\sigma$ the surface tension coefficient, $R$ the bubble's radius and $a$ the length of the edge of the initial square droplet.

\begin{figure}[bthp]
	\centering
	\subfloat[{$t / t_{\sigma} = 0.2$}] {\includegraphics[width=0.3\textwidth]{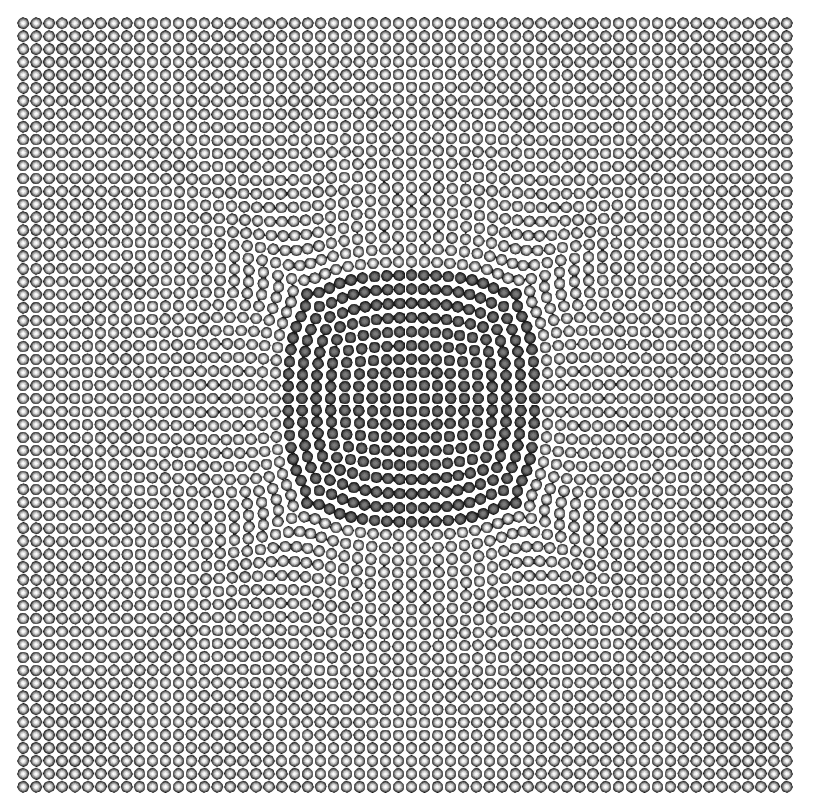}}\hspace{0.5cm}
	\subfloat[{$t / t_{\sigma} = 0.6$}] {\includegraphics[width=0.3\textwidth]{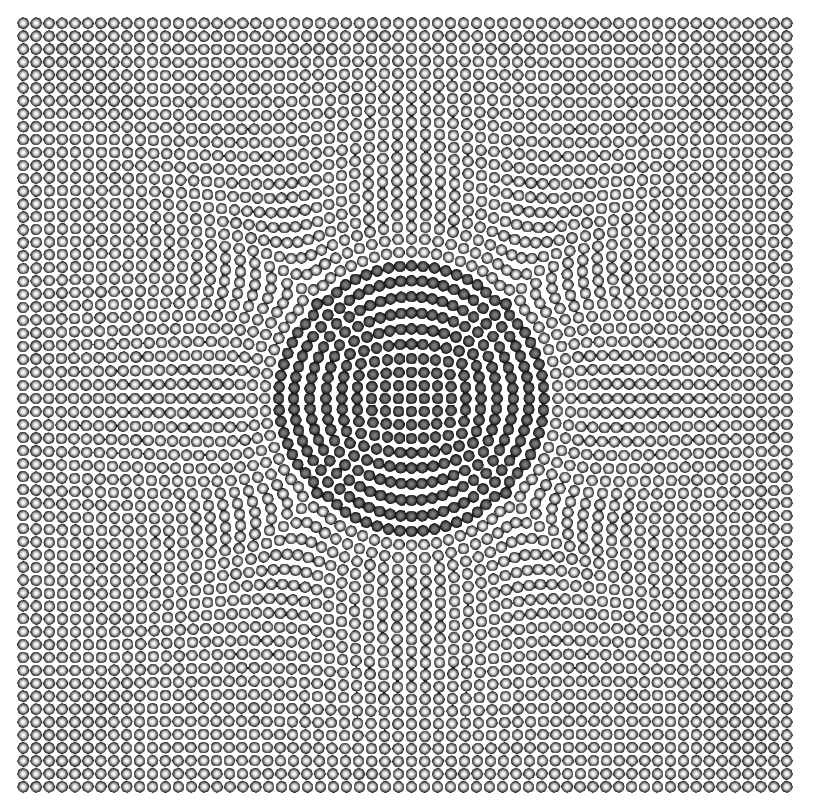}}\hspace{0.5cm}    
	\subfloat[{$t / t_{\sigma} = 1.5$}] {\includegraphics[width=0.3\textwidth]{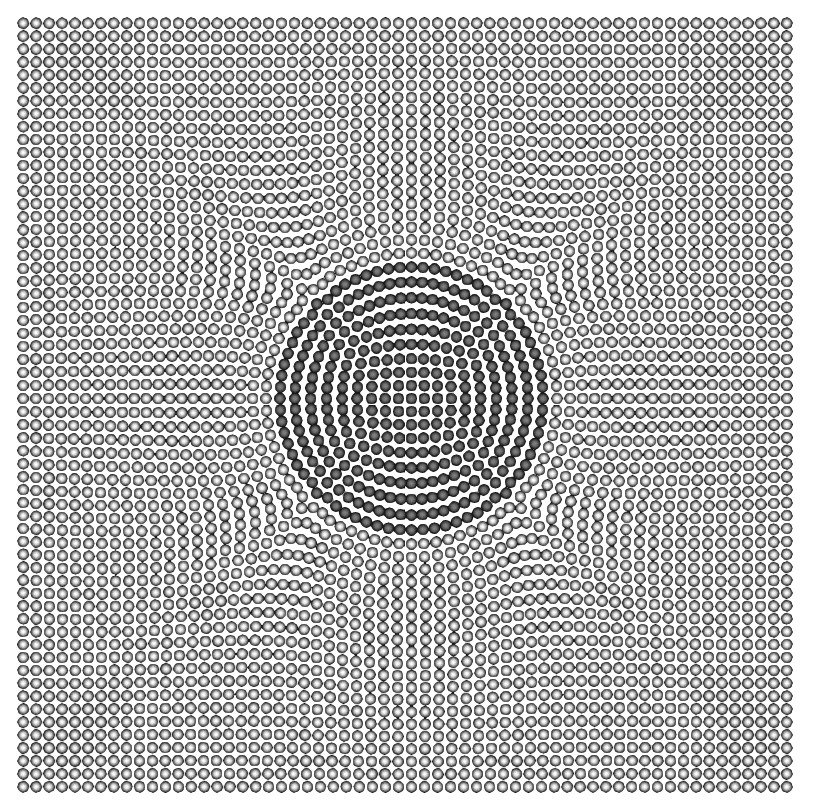}}
	\caption{Evolution of the square to droplet deformation at selected timesteps}
	\label{sqtodrop}
\end{figure}

\begin{figure}[bthp]
	\centering	
	\resizebox{0.6\textwidth}{!}{\input{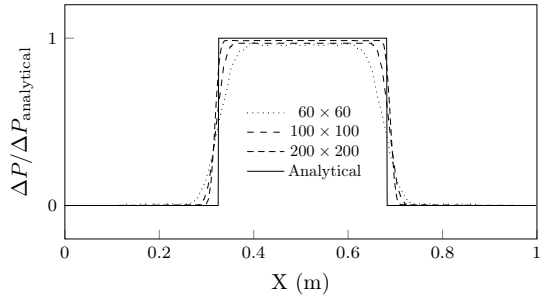}}
		\caption{Pressure profiles in the steady state}
	\label{laplace}
\end{figure}

Simulations are performed for three different resolution: $60\times 60$, $100\times 100$ and $200\times 200$ particles. The density and viscosity ratios are equal to one. The surface tension coefficient is $\sigma = 22.5~\newton\per\meter$. The whole domain is $1~\meter \times 1~\meter$ and the edge of the initial square droplet is $a=0.6~\meter$. $\gamma=7.0$ for both fluids. The time is normalized by $t_\sigma=\sqrt{\rho a^3 /\sigma}$. This test case and its parameters values are taken from~\cite{szewc2013}. On Figure~\ref{sqtodrop}, one can observe the deformation from a square to a circle in the $60\times 60$ case whereas on Figure~\ref{laplace}, pressure profiles are plotted. The error is $\leq 5\%$ at the center of the bubble for  $60\times 60$ case and $\leq 1.5\%$ for the $200\times 200$ case, but the highest resolution exhibits a steeper pressure profile in accordance with the analytical profile. Moreover, this test case allows us to confirm that the small structures (bubbles or drops made of a handful of particles) observed in Figures~\ref{flow_regime_if_corr_case1} and \ref{flow_regime_if_corr_case2} for example are not discretized enough to ensure a realistic behavior~\cite{szewc2016droplet}. The minimum number of particles for a droplet or a bubble to behave properly has yet to be exactly determine but a minimum number of $400$ particles appears necessary (with the current formulation) to recover Laplace's law with an error around $5\%$ and a reasonably steep pressure profile.

\subsection{Contact Angle}

In order to extended the validation of the surface tension model described in section~\ref{sec:st}, a test to evaluate the ability of the model to prescribe a contact angle between a wetting phase, a non-wetting phase and a solid phase is performed. Simulations are done with $60 \times 60$ particles. The density and viscosity ratios are both equal to one. $\gamma=7.0$ for both fluids. The surface tension coefficient between the wetting and non-wetting phase is $\sigma^{nw} = 100~\newton\per\meter$ whereas the one between the wetting and the solid phase is set to $\sigma^{sw}=0~\newton\per\meter$ and the one between the non-wetting and the solid phase $\sigma^{sn}$ is adjusted to prescribed the desired contact angle $\theta^{\text{prescribed}}_c$ using the Young-Laplace equation

\begin{equation}
\theta_c = \frac{\sigma^{sw} - \sigma^{sn}}{\sigma^{nw}}
\end{equation}

\begin{figure}[bthp]
	\centering
	\resizebox{0.5\textwidth}{!}{\input{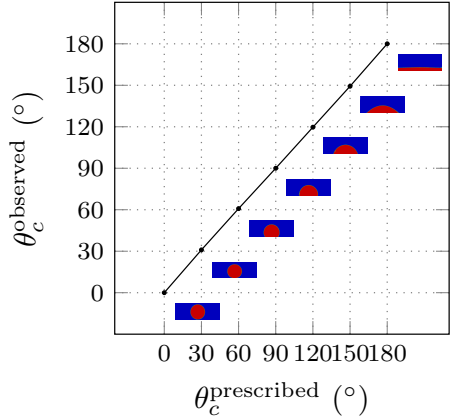}}
	\caption{Different measured contact angles}
	\label{cangle}
\end{figure}

The whole domain is $1~\meter \times 1~\meter$ and the edge of the initial rectangle droplet is $0.6~\meter \times 0.3~\meter$. At steady state, the observed contact angle $\theta^{\text{observed}}_c$ is measured and reported in Figure~\ref{cangle}. The coefficient of determination is $0.9894$ confirming that the model can accurately reproduce a prescribed contact angle.

\subsection{Capillary Rise}

\begin{figure}[bthp]
	\centering
	\resizebox{0.8\textwidth}{!}{\input{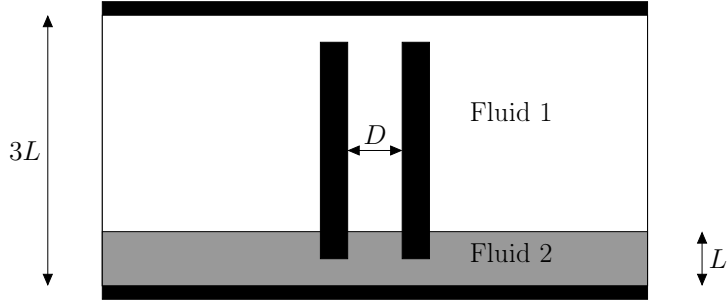}}
	\caption{The capillary rise problem}
	\label{caprise_case}
\end{figure}

The well-known capillary rise problem is a simple test case to further verify the ability of the model to reproduce contact line dynamics. The problem is described in Figure~\ref{caprise_case}. It consists of two fluids, one on top of the other. Two vertical parallel plates are immersed in the fluids. Thanks to the action of surface tension, the lower fluid will rise along the parallel plates forming a meniscus between them. The lower liquid height reached at steady state follows Jurin's law

\begin{equation}
	H_{\text{J}} = \frac{2 \sigma^{nw} \cos{\theta_c}}{(\rho_w - \rho_n) g D}
\end{equation}

\noindent with $\sigma^{nw}$ the surface tension coefficient between the wetting and the non-wetting phase, $\theta_c$ the contact angle, $\rho_w$ and $\rho_n$ the densities of the wetting and non-wetting phase, $g$ the gravity and $D$ the horizontal distance between the two plates.

\begin{figure}[bthp]
	\centering
	\includegraphics[width=0.8\textwidth]{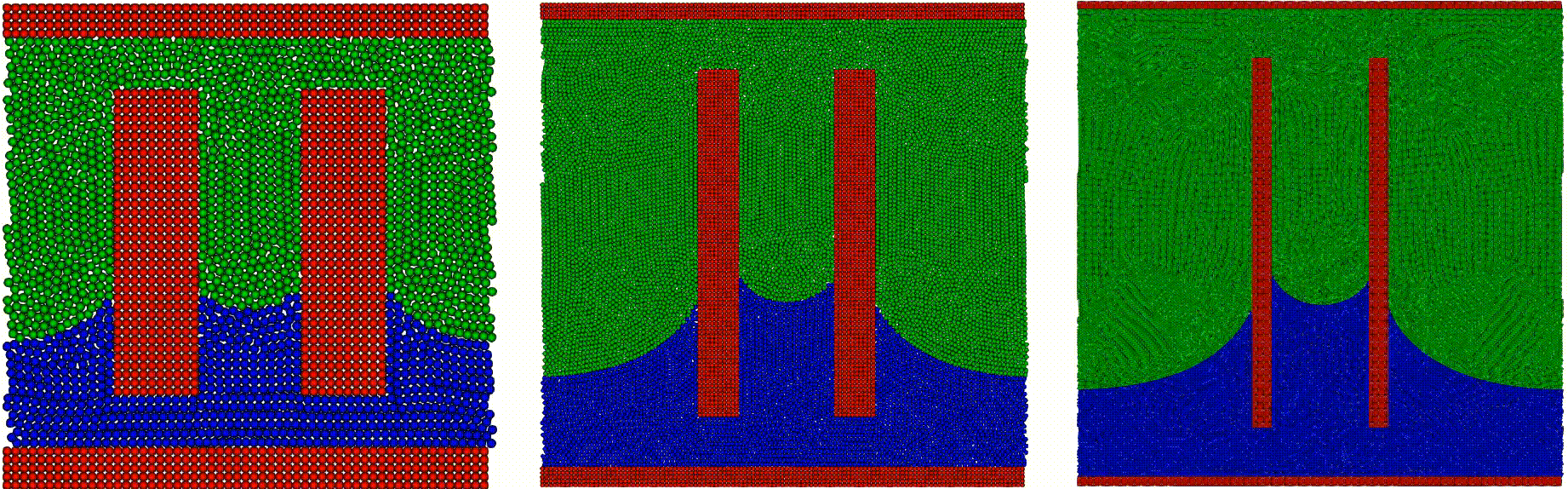}
	\caption {Steady state results with, from left to right, $60 \times 60$, $140 \times 140$ and $223 \times 223$ particles for $\theta_c=30^{\circ}$}
	\label{caprise_pics}
\end{figure}

\begin{figure}[bthp]
	\centering
	\resizebox{0.6\textwidth}{!}{\input{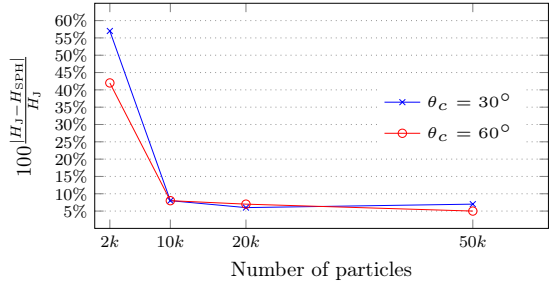}}
	\caption {Measured error between the final height of the lower fluid for different resolutions and different contact angles}
	\label{caprise_err}
\end{figure}

Simulations were performed with four different resolutions : $60 \times 60$, $100\times 100$, $140\times 140$ and $223\times 223$ particles and two different contact angles. The surface tension coefficient is set to $1.88~\newton\per\meter$. The initial height of the lower fluid is $L=0.33~\meter$. The density and viscosity ratios are both equal to one. $\gamma=7.0$ for both fluids. The final height of the lower fluid is then measured and reported in Figure~\ref{caprise_err}. On Figure~\ref{caprise_pics}, one can observe the final particle distributions for different resolutions. The top and bottom boundaries are modeled with non-slip conditions whereas the left and right boundaries are periodic.

The lowest resolution is not able to correctly reproduce the fluid rise. Moreover, because of the low resolution, the thickness of the parallel plates is too important and does not properly represent the expected geometry. For the other resolutions, the error is between $5\%$ and $10\%$ showing the ability of the model to reproduce the expected behavior with a reasonable accuracy.

\subsection{Rayleigh-Taylor Instability}

The Rayleigh-Taylor instability is a famous two-phase instability where a heavy fluid is place on top of a light fluid with a given interface shape and submitted to gravity. Several SPH researchers have reproduced this case with SPH~\cite{grenier2009,szewc2013}. The test case and its parameters are borrowed from~\cite{grenier2009}. The computational domain is twice as high as long, $H\times L$ with $H=2L$ and populated with $40000$ particles. The density ratio is $1.8$ while the viscosity ratio is $1$. Gravity is set $g=9.81~\meter\per\second^{-2}$ and oriented downwards. Therefore, the viscosity $\nu$ is adjusted to math the desired Reynolds number $Re = \sqrt{\frac{(H/2)^3 g}{\nu}} = 420$. $\gamma=7.0$ for both fluids. No surface tension is used. No slip boundary conditions are applied at the walls. The interface is initialized as follows : $y= 1-\sin(2\pi x)$. Time $t$ is nondimensionalized by $t_g=1/\sqrt{g/H}$.

\begin{figure}[bthp]
	\centering
	\includegraphics[width=0.6\textwidth]{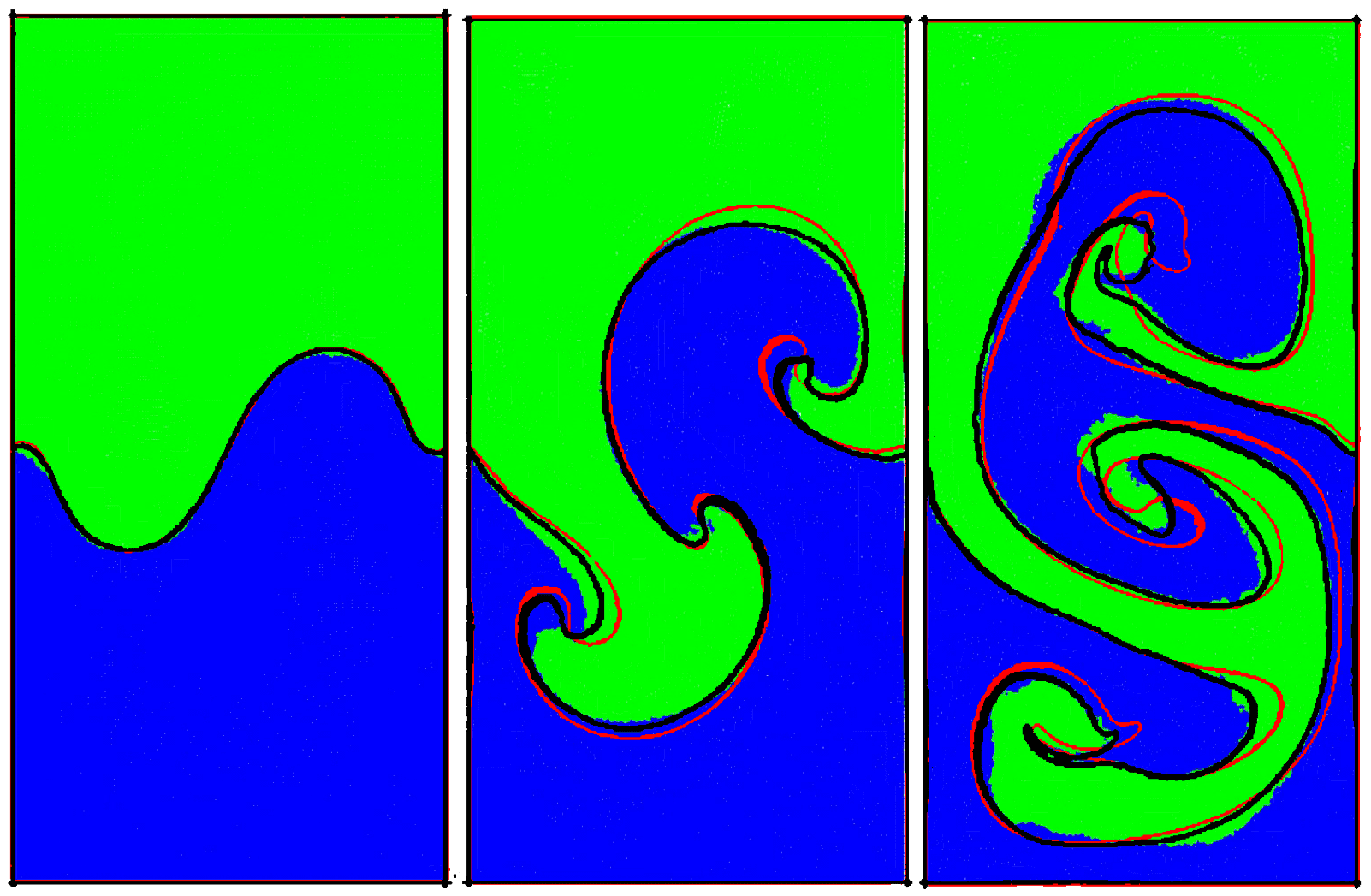}
	\caption {SPH phases distribution of Rayleigh-Taylor instability at selected timesteps : $t/t_g=1$, $3$ and $5$. Superposed with SPH interface (in black) and with Level-Set interface (in red) both extracted from~\cite{grenier2009}}
	\label{rt_vorta}
\end{figure}

\begin{figure}[bthp]
	\centering
	\subfloat[Current work\label{rt_vortb}] {\includegraphics[width=0.58\textwidth]{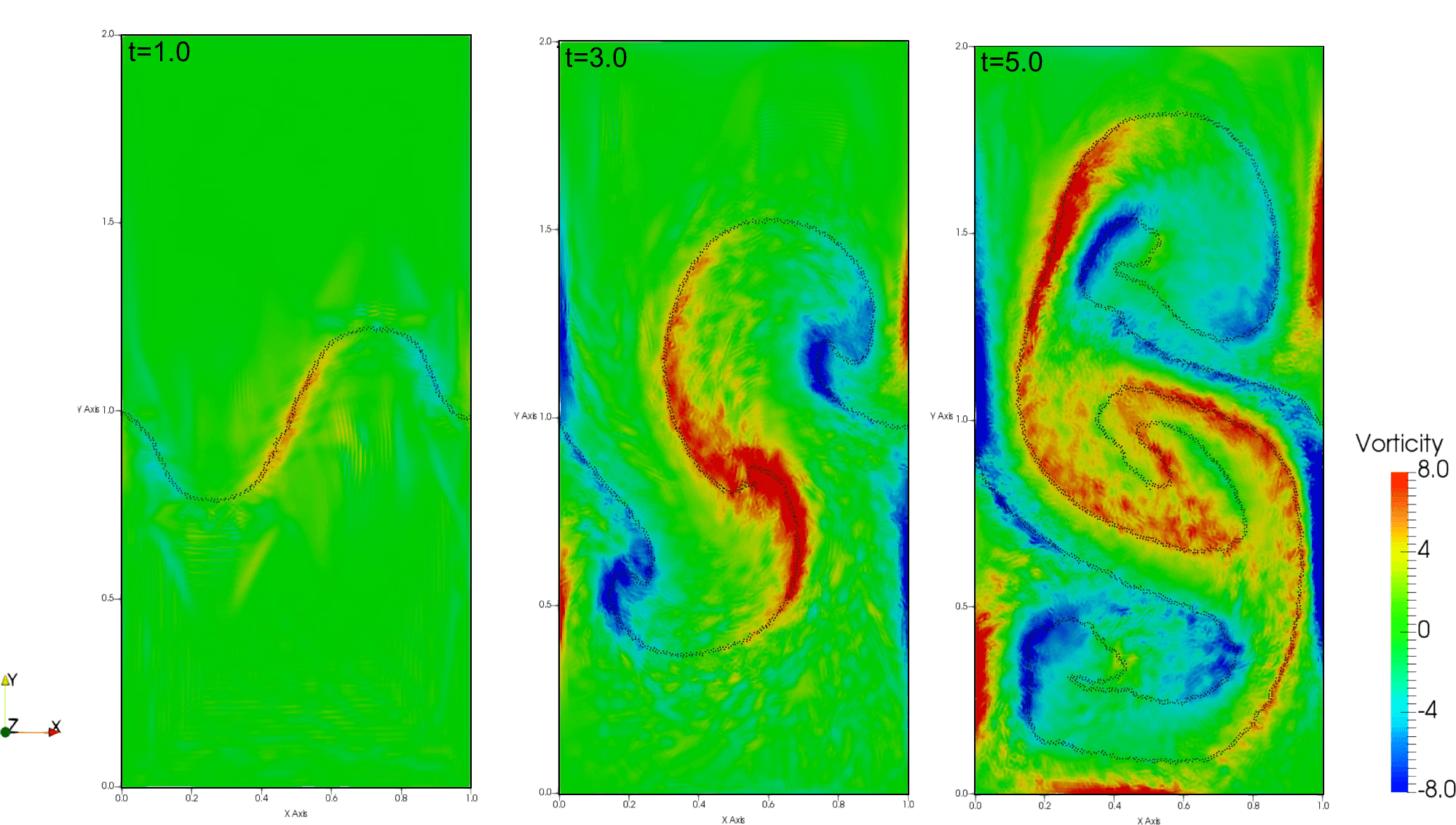}}\\
		\subfloat[SPH (from~\cite{grenier2009})\label{rt_vortc}] {\includegraphics[width=0.6\textwidth]{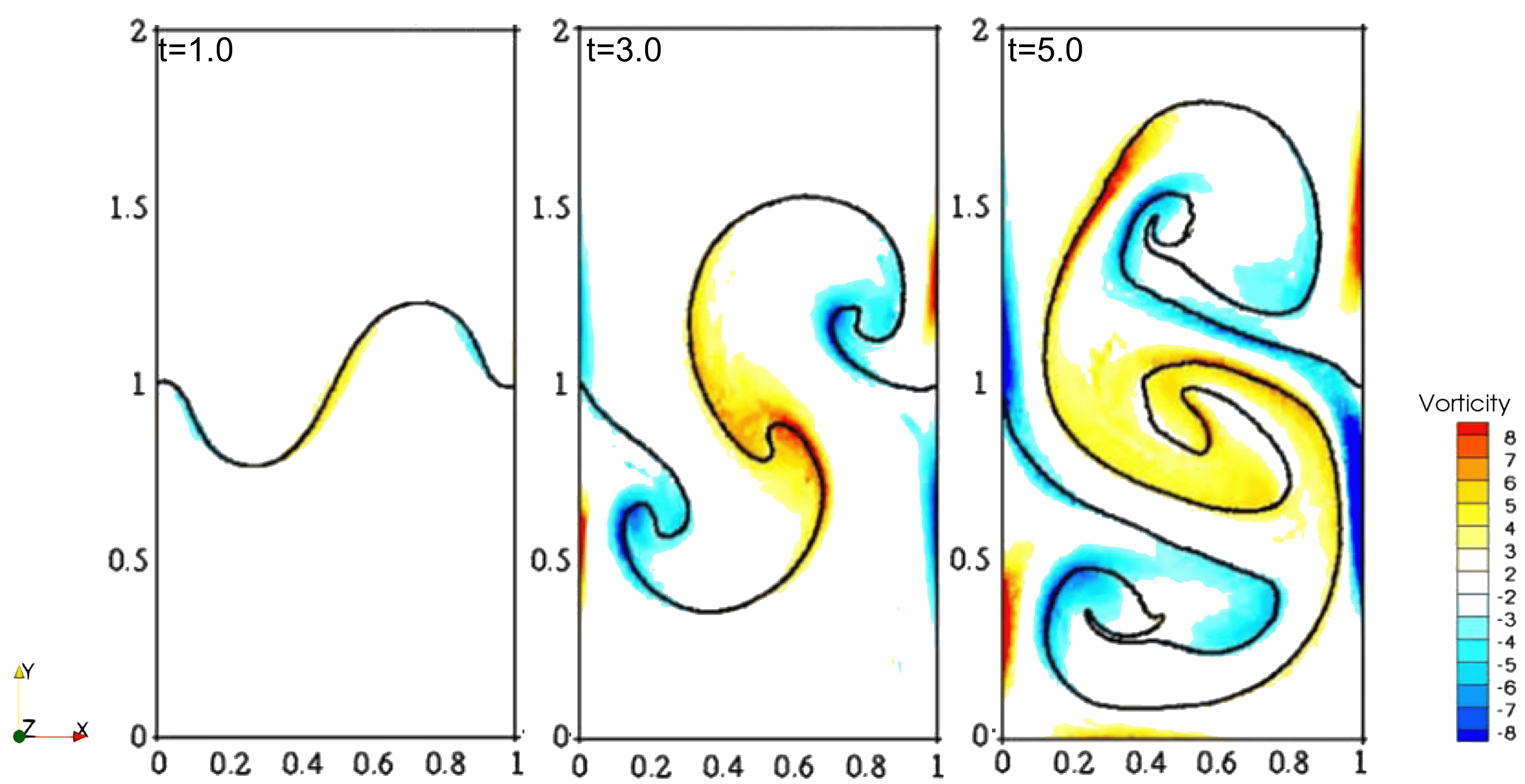}}\\
			\subfloat[Level Set (from~\cite{grenier2009})\label{rt_vortd}] {\includegraphics[width=0.6\textwidth]{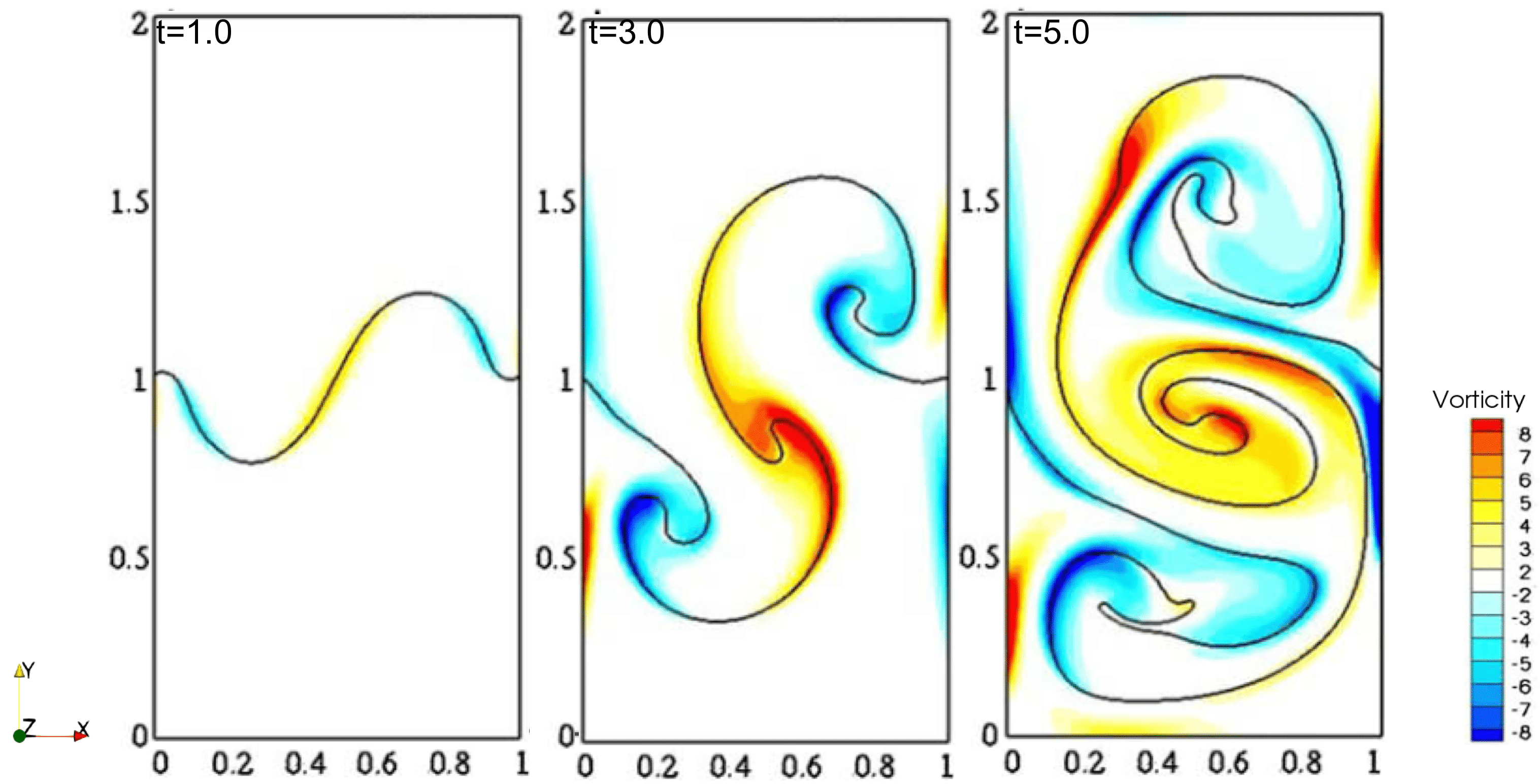}}
	\caption {Vorticity results of Rayleigh-Taylor instability at selected timesteps : $t/t_g=1$, $3$ and $5$}
	\label{rt_vort}
\end{figure}

The distribution of the two phases is shown at selected timesteps on Figure~\ref{rt_vorta} superposed with results from~\cite{grenier2009}. In order to validate the model in a more quantitative way than just looking at the shape of the interface, vorticity has been calculated and is displayed on Figure~\ref{rt_vortb} at selected timesteps. These vorticity plots can be compared with results presented on Figures 5 and 6 in~\cite{grenier2009} (and reproduced here on Figures~\ref{rt_vortc} and~\ref{rt_vortd} with permission) showing good agreement. Note that the SPH results from~\cite{grenier2009} where done with $180000$ particles.

\end{document}